\documentclass{pasa}%

\usepackage{graphicx}

\title[Giant Radio Galaxy PKS 2250$-$351 in Abell 3936]{PKS 2250$-$351: A Giant Radio Galaxy in Abell 3936}
\author[Seymour, N. et al.]{Seymour N.$^1$, Huynh M.$^{2,3}$, Shabala S.S.$^4$, Rogers J.$^4$, Davies L.J.M.$^3$, 
Turner R.J.$^4$, O'Brien A.$^5$, Ishwara-Chandra C.H.$^6$, Thorne J.E.$^3$, Galvin T.J.$^2$, Jarrett T.$^7$, Andernach H.$^8$, Anderson C.$^2$, Bunton J.$^5$, Chow K.$^5$, Collier J.D.$^{9,10}$, Driver S.$^3$, Filipovic M.$^9$, G\"urkan G.$^2$, Hopkins A.$^{11}$, Kapi\'nska A.D.$^{12}$, Leahy D.A.$^{13}$, Marvil J.$^{12}$, Manojlovic P.$^{10,5}$, Norris R.P.$^{10,5}$, Phillips. C.$^5$,  Robotham A.$^3$, Rudnick L.$^{14}$, Singh V.$^{15}$, and White S.V.$^{1,16}$
\affil{$^1$International Centre for Radio Astronomy Research, Curtin University, Bentley, WA 6102, Australia}
\affil{$^2$CSIRO Astronomy and Space Science, 26 Dick Perry Avenue, Kensington, WA 6151, Australia}%
\affil{$^3$International Centre for Radio Astronomy Research, M468, University of Western Australia, Crawley, WA 6009, Australia}%
\affil{$^4$School of Natural Sciences, University of Tasmania, Private Bag 37, Hobart, TAS 7001, Australia}
\affil{$^5$CSIRO Astronomy and Space Science, PO Box 76, 1710, Epping, NSW, Australia}%
\affil{$^6$National Centre for Radio Astrophysics, TIFR, Post Bag No. 3, Ganeshkhind Post, 411007 Pune, India}%
\affil{$^7$Astronomy Department, University of Cape Town, Private Bag X3, Rondebosch 7701, South Africa}
\affil{$^8$Depto. de Astronom\'ia, DCNE, Universidad de Guanajuato, Apdo. Postal 144, Guanajuato, CP 36000, Gto., Mexico}
\affil{$^9$The Inter-University Institute for Data Intensive Astronomy (IDIA), Department of Astronomy, University of Cape Town, Rondebosch 7701, South Africa }
\affil{$^{10}$School of Computing, Engineering and Mathematics, Western Sydney University, Locked Bag 1797, Penrith, NSW 2751, Australia}
\affil{$^{11}$Australian Astronomical Optics, AAO-Macquarie, Faculty of Science and Engineering, Macquarie University, 105 Delhi Rd, North Ryde, NSW 2113, Australia}
\affil{$^{12}$National Radio Astronomy Observatory, 1003 Lopezville Road, Socorro, NM 87801, USA}
\affil{$^{13}$Department of Physics and Astronomy, University of Calgary, Calgary, Alberta, T2N 1N4, Canada}
\affil{$^{14}$Minnesota Institute for Astrophysics, School of Physics and Astronomy, University of Minnesota, 116 Church Street SE, Minneapolis, MN 55455, USA}
\affil{$^{15}$Astronomy and Astrophysics Division, Physical Research Laboratory, Ahmedabad 380009, India}
\affil{$^{16}$Department of Physics and Electronics, Rhodes University, PO Box 94, 6140 Grahamstown, South Africa}
}

\jid{PASA}
\doi{10.1017/pas.\the\year.xxx}
\jyear{\the\year}

\usepackage{aas_macros}
\usepackage{hyperref} 
\hypersetup{colorlinks,citecolor=blue,linkcolor=blue,urlcolor=blue}

\begin{document}

\begin{frontmatter}
\maketitle

\begin{abstract}

We present a detailed analysis of the
radio galaxy PKS 2250$-$351, a giant of 1.2\,Mpc projected size, its host 
galaxy, and its environment. 
We use radio data from the Murchison Widefield Array, 
the upgraded Giant Metre-wavelength Radio Telescope, the Australian 
Square Kilometre Array  Pathfinder, and the Australia Telescope 
Compact Array to model the jet power and age. 
Optical  and infra-red data come from the Galaxy And Mass Assembly 
 (GAMA) survey and provide information on the host galaxy  and environment. 
 GAMA spectroscopy confirms that PKS 2250$-$351 lies at $z=0.2115$ in the irregular,  and likely unrelaxed, 
cluster Abell 3936. 
We find its host is a massive, `red and dead' elliptical  galaxy
with negligible star formation  but with a 
highly obscured active galactic  nucleus dominating the 
mid-infrared emission. Assuming it lies on the local $M-\sigma$ 
relation it has an Eddington accretion rate of 
$\lambda_{\rm EDD}\sim 0.014$.
 We find that the lobe-derived jet power  (a time-averaged 
measure) is an order of magnitude greater than
the hotspot-derived jet power (an instantaneous measure).
We propose that over the lifetime of the 
observed radio emission ($\sim 300\,$Myr) the accretion has switched 
from an inefficient 
advection dominated mode to a thin-disc efficient mode, consistent with the decrease in jet power.
 We also suggest that the asymmetric radio morphology is due to its 
environment, with the host of PKS 2250$-$351 lying to the west of 
the densest concentration of galaxies in Abell 3936.

\end{abstract}

\begin{keywords}
radio continuum: galaxies  -- galaxies: active 
\end{keywords}
\end{frontmatter}

\section{INTRODUCTION }
\label{sec:intro}

Radio-loud active galactic nuclei (AGN) are one manifestation of 
the super-massive black holes (SMBH, $10^5 - 10^9\,$M$_\odot$)
which lie at the centre of nearly all galaxies. They are powered by 
bipolar relativistic outflows (i.e. jets) of ionised material 
originating close to the SMBH \citep{Rees:78}. In the local Universe, 
radio-loud AGN are typically hosted by passive elliptical galaxies 
with negligible accretion rates 
 \citep[Eddington ratio $<1\%$, see][and below]{Heckman:14}. However, 
many of the most powerful radio-loud AGN in the distant Universe 
are powered by SMBH with very high accretion rates 
\citep[e.g.][]{Drouart:14}. This difference is thought to be due to 
differing natures of the accretion disc at high and low accretion 
rates. Typically these rates are normalised to the theoretical maximum 
accretion rate (the `Eddington' rate, $\lambda_{\rm EDD}=1$). 
By comparison to stellar mass
black holes in our own galaxy \citep{Merloni:03,Fender:04,McHardy:06},
SMBH with low Eddington accretion rates, $\lambda_{\rm EDD}\le 0.01$,  
are thought to have a thick  disc, efficient at producing jets whereas 
SMBH with high Eddington accretion rates, $\lambda_{\rm EDD}> 0.01$, 
are thought to have a thin  disc which is less efficient at producing 
jets. However, both states of accretion  disc can produce jets with 
powers proportional to the SMBH mass, spin, and accretion 
\citep{Meier:02}. However, converting from the observed radio luminosity 
to jet power is complex and involves taking into account the environment 
of the AGN \citep[e.g.][]{Krause:19}.

Giant radio galaxies (GRG) are a rare class of radio-loud AGN with 
projected maximum angular extents $>1\,$Mpc of which only a few hundred
are known \citep[][and references therein]{Kuzmicz:18}.
It is unusual for radio sources to grow so large as it presumably requires a low enough density
in the local intergalactic medium (IGM) to travel far, but a high enough density
that the observed jets have something to work against, 
thereby creating ` hotspots' where the jets terminate against the IGM.
Giant radio sources are believed to be  the late stage in the evolution of otherwise normal radio galaxies such as Cygnus A \citep[]{Ishwar:99}. If this is true then 
there must be many more giant radio sources than known today, but are 
missed due to observational selection effects. \cite{Komberg:09} have shown that
GRGs are no different to the population of less extended radio-loud AGN
in terms of lobe asymmetry, prevalence of high 
excitation lines in host, radio powers, and environments. Indeed, 
GRGs are found in both isolated environments and rich clusters.

Broadband radio surveys can provide key insights into the 
astrophysics occuring at these wavelengths
in radio-loud AGN \citep[e.g.][]{Callingham:17}.
To model the radio emission and constrain radio jet powers and ages 
 \citep[e.g.][]{Shabala:08,Turner:15,Hardcastle:19,Turner:18a} 
requires surveys with broad-band radio data. 
 The ideal region of sky for this work
is the Galaxy And Mass Assembly \citep[GAMA,][]{Driver:09}
survey field at RA$\sim 23^h$ and dec.$\sim -32.5^\circ$ (hereafter the `G23 
field') due to its declination and superb multi-wavelength coverage. 
As part of this effort, the GAMA 
Legacy  Australia Telescope 
Compact Array (ATCA) Southern Survey (GLASS) is providing 
deep ($\sim 30\,\mu$Jy/beam RMS) radio observations at 5.5 and 9.5\,GHz over G23.
As an extension of  GLASS  this field has also been observed with the 
upgraded Giant Metre-wavelength Radio Telescope 
(uGMRT) as part of the uGMRT/GLASS project at 
$250-500$ and $500-800\,$MHz.

\begin{figure*}[th]
\begin{center}
\includegraphics[angle=0,width=1.5\columnwidth]{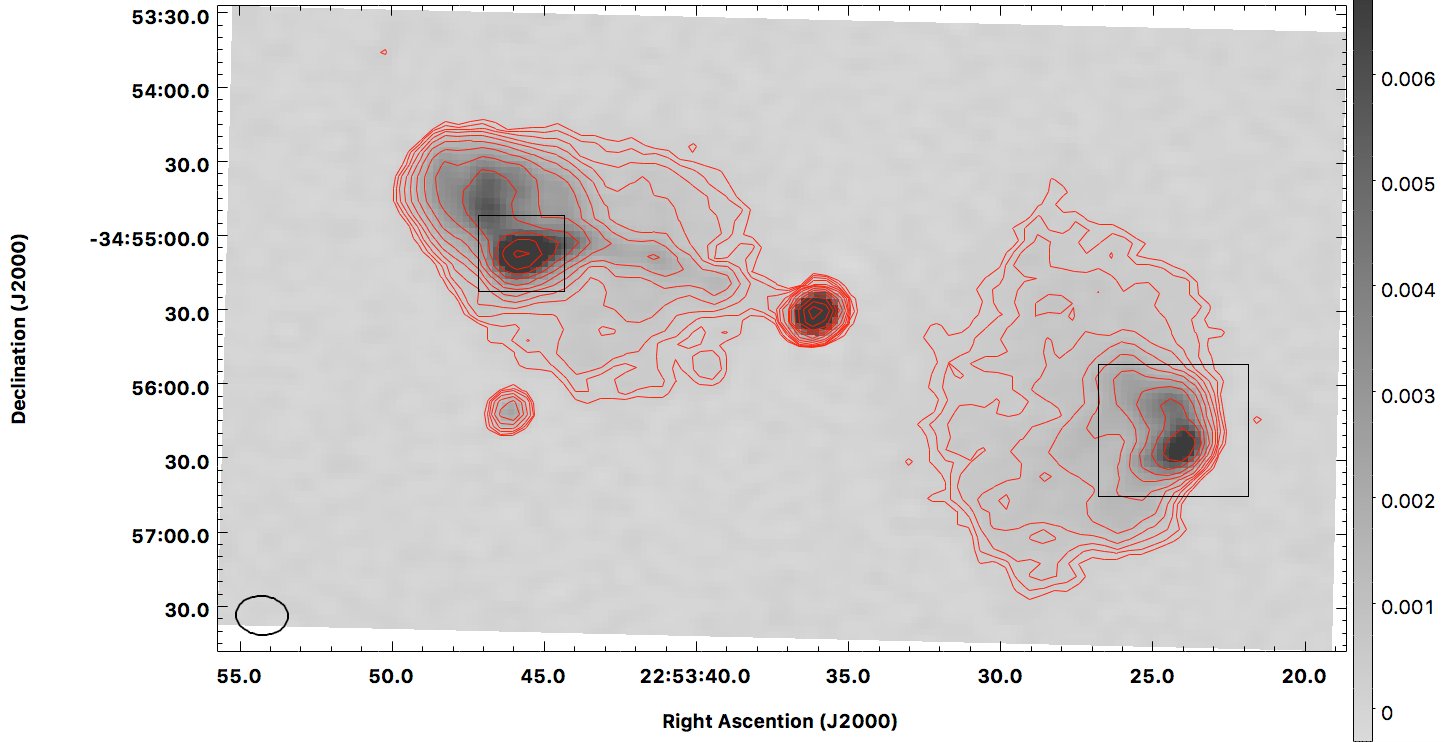}
\caption{Greyscale image of PKS 2250$-$351 from our 888\,MHz ASKAP continuum data. The local 
rms is $\sim 64\,\mu$Jy/beam and the restoring beam has a FWHM of $10.55'' \times 7.82''$ (indicated 
in the lower left). 
 The greyscale is a linear stretch  in Jy/beam as indicated by the colourbar. The red contours start at $4\sigma$ and increase by factors of $\sqrt{2}$.
The two  black rectangles indicate the regions of the GLASS data presented
in Figure~\ref{fig:glass}.
The core is clearly identified as well as  hotspots positioned on top of diffuse emission from the lobes. Within the eastern lobe the jet is observed with several knots.
The total  angular size is $5.66'$
($\equiv\,1.17\,$Mpc$^1$)  with equal lobe lengths. However the width of each 
lobe is markedly different with the western lobe being $1.4\times$ 
wider than the eastern lobe.}

\label{fig:askap}
\end{center}
\end{figure*}

In this paper we present the most sensitive radio images of PKS 
2250$-$351,  a GRG lying in G23, using  Australian 
Square Kilometre Array Pathfinder (ASKAP),
uGMRT, ATCA, and 
the Murchison Widefield Array (MWA).
These observations allow us to detect various morphological features, in particular, 
the diffuse extended emission in the lobes  and hotspots.
We study the broad-band radio properties of PKS 2250$-$351, the multi-wavelength properties of the its host galaxy, its AGN characteristics, 
and environment.
This paper is organised into the following sections. Section~\ref{sec:obs} presents the 
radio observations and GAMA data on the host galaxy. 
In Section~\ref{sec:modelling} we present 
the modelling and analysis of both the radio data and the host galaxy data. 
We discuss these results in Section~\ref{sec:dis}  and conclude this work in Section~\ref{sec:con}. 
Throughout this paper
we use a flat `concordance' cosmology of $H_0=70\,$km\,s$^{-1}$ and 
$\Omega_M = 0.3 = 1- \Omega_\Lambda$. We define radio spectral index, $\alpha$, by $S_\nu\propto\nu^\alpha$.

\section{Observations}
\label{sec:obs}

\subsection{Radio Data}
\label{sec:raddat}

\cite{Brown:91} associated 
PKS 2250$-$351 with the cluster Abell 3936 and assigned 
 both an estimated redshift from \cite{Abell:89}.
Note all the radio surveys mentioned below (literature and new) 
use, or are matched to within 
$1-2\%$ of, the \cite{Baars:77} flux density scale.

\begin{figure*}[t]
\begin{center}
\includegraphics[angle=0,width=1.6\columnwidth]{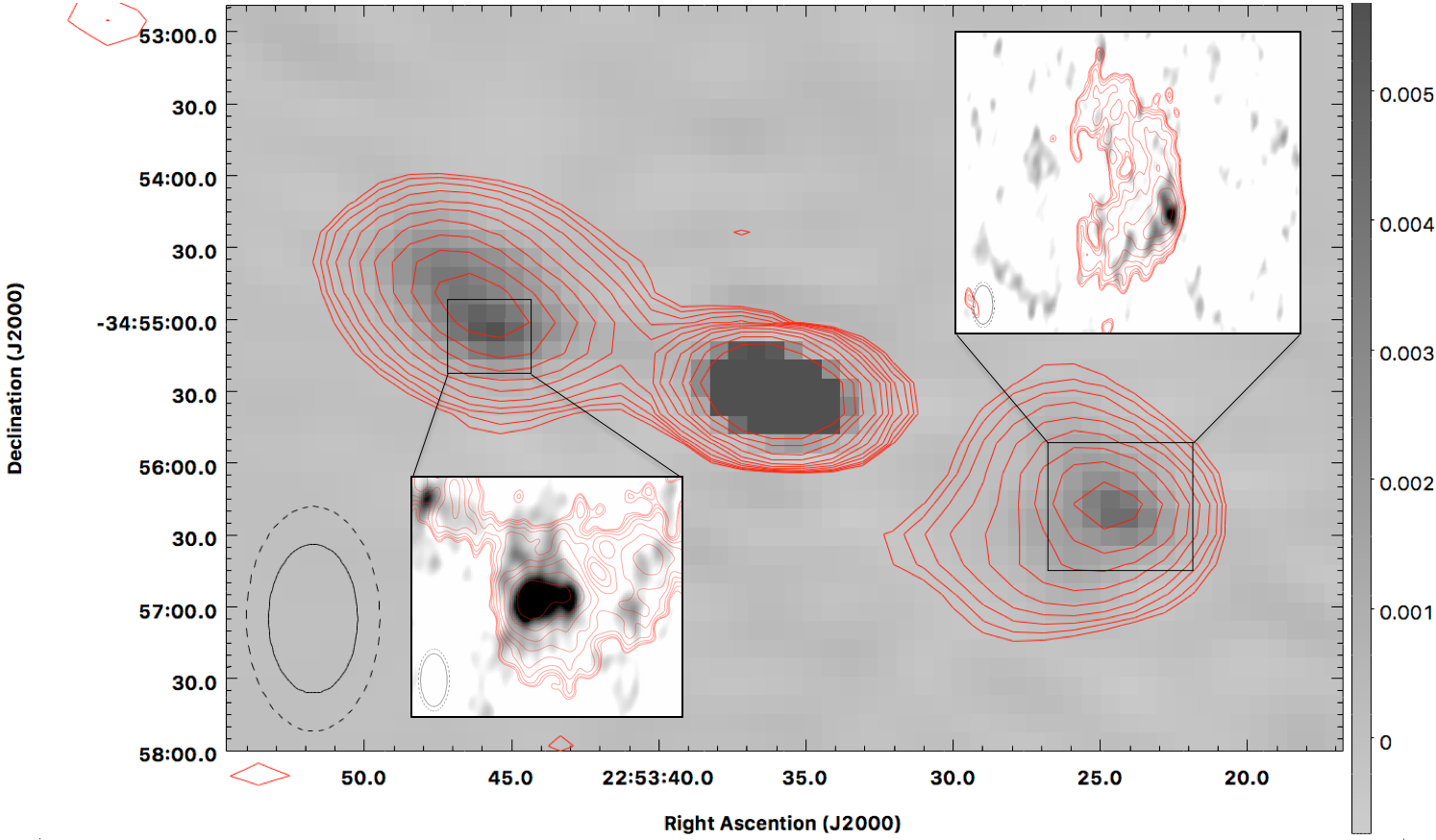}
\caption{Greyscale ATCA 9.5\,GHz images with 5.5\,GHz images overlaid as red contours. 
The contours of the main panel (from the `green time' data)
start at $3\times\sigma$ ($\sigma=200\,\mu$Jy/beam). 
The contours of the inset panels (close-ups of the hotspots from GLASS) start at $3\times\sigma$ 
($\sigma=24\,\mu$Jy/beam). Both sets of contours increase by factors of $\sqrt{2}$. 
The resolution is 
indicated in the lower left of each panel by the ellipses (solid 9.5\,GHz and dashed 5.5\,GHz). 
The greyscale  stretch of the main figure, indicated by the colour-bar, 
is in Jy/beam. The stretch of the inserts is a linear stretch from $-0.2$ to $+0.3\,$mJy/beam.}

\label{fig:glass}
\end{center}
\end{figure*}

\subsubsection{Literature Radio Data}
\label{sec:radlit}
The NRAO VLA Sky Survey \citep[NVSS at 1.4\,GHz,][]{Condon:98} and the Sydney 
University Molonglo Sky Survey \citep[SUMSS at 0.843\,GHz,][]{Bock:99} images of PKS 2250$-$351 reveal the classic lobe-core-lobe morphology, 
typical of a resolved radio galaxy. From these images we observe that 
it is approximately $5'$ in maximum extent. Based upon  the 
spectroscopic redshift of  its host, $z=0.2115$, 
\citep[from the 2dfGRS][see Section~\ref{sec:anclit}]{Colless:01} 
this  radio source
is  $\ge1\,$Mpc\footnote{using 3.45\,kpc/$''$ from\\ 
{\tt astro.ucla.edu/$\sim$wright/CosmoCalc.html} \citep{Wright:06}} 
in size.
The Tata Institute of Fundamental Research (TIFR) GMRT Sky Survey Alternative Data Release 1 \citep[TGSS ADR1 at $\sim 0.15\,$GHz,][]{Intema:17}  
detects the lobes,  and presents faint, uncatalogued emission 
at the position of the core at $15-20\,$mJy/beam.
The core of PKS 2250$-$351 is detected in the  Australia Telescope 
20\,GHz (AT20G) survey \citep{Murphy:06} which also provides 5 and 8\,GHz 
measurements. In the AT20G catalogue this source  has a quality flag of `poor', 
meaning the flux densities were measured from lower quality data.
The radio photometry from 
NVSS, SUMSS, AT20G and TGSS  is presented in 
Table~\ref{tab:radio}.

We also use data from the low-frequency  GaLactic and Extragalactic All-sky
MWA (GLEAM) survey \citep{wayth:15} conducted in phase 1 of the 
MWA \citep{Tingay:13}. The first GLEAM extragalactic data release \citep[hereafter DR1][]{NHW:17} provides 20 band photometry across $70-230\,$MHz
over a large fraction of the southern sky. 
Radio sources were detected, and had their flux densities measured, in a deep, high resolution $170-230\,$MHz image. Each detected source then had its flux density measured in each of the 20 sub-bands using the positions found in the deep $170-230\,$MHz image as priors \citep[see][for more details]{NHW:17}. This approach ameliorated the blending of sources at lower frequencies.
The two lobes of PKS 2250$-$351 are detected at high significance at $170-230\,$MHz
and their published MWA flux densities are reported in Table~\ref{tab:radio}.  Note the corresponding uncertainties have 
had an $8\%$ systematic uncertainty added in quadrature.

\subsubsection{ASKAP Early Science Data}

ASKAP \citep{Johnston:07,McConnell:16} utilises the 
revolutionary phased array feeds (PAFs) that densely sample the 
focal plane of the 12\,m
antennas with 188 dipoles. To cover the full field of view of the PAF 36 beams are electronically formed by 
combining the signals from individual dipoles. 
 Each beam is correlated with the corresponding beam from every other antenna.
These data were obtained before the fringes were tracked for each beam 
(rather than just the centre of the field of view) which results in 
significant peak flux density suppression for sources near the 
field edges due to bandwidth smearing.

The G23 field was observed several times during the commissioning and early science period of ASKAP  \citep[e.g.][]{Leahy:19}. In 
this paper we include data observed on $30^{\rm th}$ September 2018 with 28 of the final 36 antennas  which simultaneously observed 36 primary beams arranged in a square $6\times6$ footprint as part of  EMU \citep[Evolutionary Map of the Universe][]{Norris:11} project early science.  
 The total integration time was 11\,hours with 288\,MHz bandwidth centred on 888\,MHz.

 The data were processed using the {\tt ASKAPsoft} pipeline\footnote{\url{https://www.atnf.csiro.au/computing/software/askapsoft/sdp/docs/current/pipelines/introduction.html}} on the Galaxy supercomputer hosted by the Pawsey Supercomputing Centre. The data for each of the 36 beams were bandpass and flux calibrated by observing the primary calibrator source PKS B1934$-$638 at the centre of each primary beam for approximately three minutes. The bandpass was then solved using the \cite{Reynolds_1994} model and the solutions were applied to the science target observations. Images for each beam were produced independently in parallel by gridding the visibility data using the {\tt WProject ASKAPsoft} gridder then deconvolving the dirty images with the {\tt BasisfunctionMFS} clean solver, an improved version of the {\tt MultiScale} algorithm available in {\tt CASA} \citep{McMullin:07}. Two iterations of phase-only self-calibration were performed: the first used only a single delta function scale and included only detected components at $>20\sigma$ in the calibration model. The second iteration also used a single delta function scale, but with a lower component threshold of $8\sigma$. The self-calibration solutions were then applied to the data and a final image was produced for each beam using three deconvolution scales: 0 (i.e. the delta function), 15, and 30 pixels. All 36 individual beam images were then combined using a linear mosaic algorithm that corrects for the primary beam attenuation and combines the images using a weighted average.

In Figure~\ref{fig:askap} we present the continuum image of  a cutout around 
PKS 2250$-$351  from these data. This image  nicely reveals 
the bright  hotspots located on top of diffuse lobe emission and it also highlights the eastern jet. 
The high fidelity of the image is due to the superb uv-coverage 
 from a long integration with 36 antennas of ASKAP\footnote{\tt 
www.atnf.csiro.au/projects/askap/config.html}.
The restoring beam used in this image was $10.55''\times7.82''$ 
(full width half maximum for an elliptical Gaussian) with a 
beam position angle (BPA) of $86.8^\circ$.

Since PKS 2250$-$351 lies near the field 
edge, we do not use the flux density for the core as it is highly smeared. However, as flux  
is conserved, we can measure the total flux of each lobe by summing the flux
density in bespoke irregular polygons around each lobe  (tracing the approximate $3\sigma$ contour and with no sigma-clipping 
applied). 
We convert from the image
native units of Jy/beam to mJy using the beam size\footnote{The area of the beam is defined as 
$\Omega_{\rm beam}=\frac{\pi\,\theta_{\rm maj}\times\theta_{\rm min}}{4\,ln(2)}$.}  
converted to square pixels. Uncertainties are derived from the RMS measured within
these polygons multiplied by the square root of the area of the polygon in units
of beam size. These flux densities and uncertainties 
are reported in Table~\ref{tab:radio}.

From the image in  Figure~\ref{fig:askap} we can more accurately estimate the 
total size to be  $5.66'$ ($\equiv\,1.17\,$Mpc$^1$, projected 
size) and find the lobe lengths to be equal.

\begin{figure*}[t]
\begin{center}
\includegraphics[angle=0,width=1.5\columnwidth]{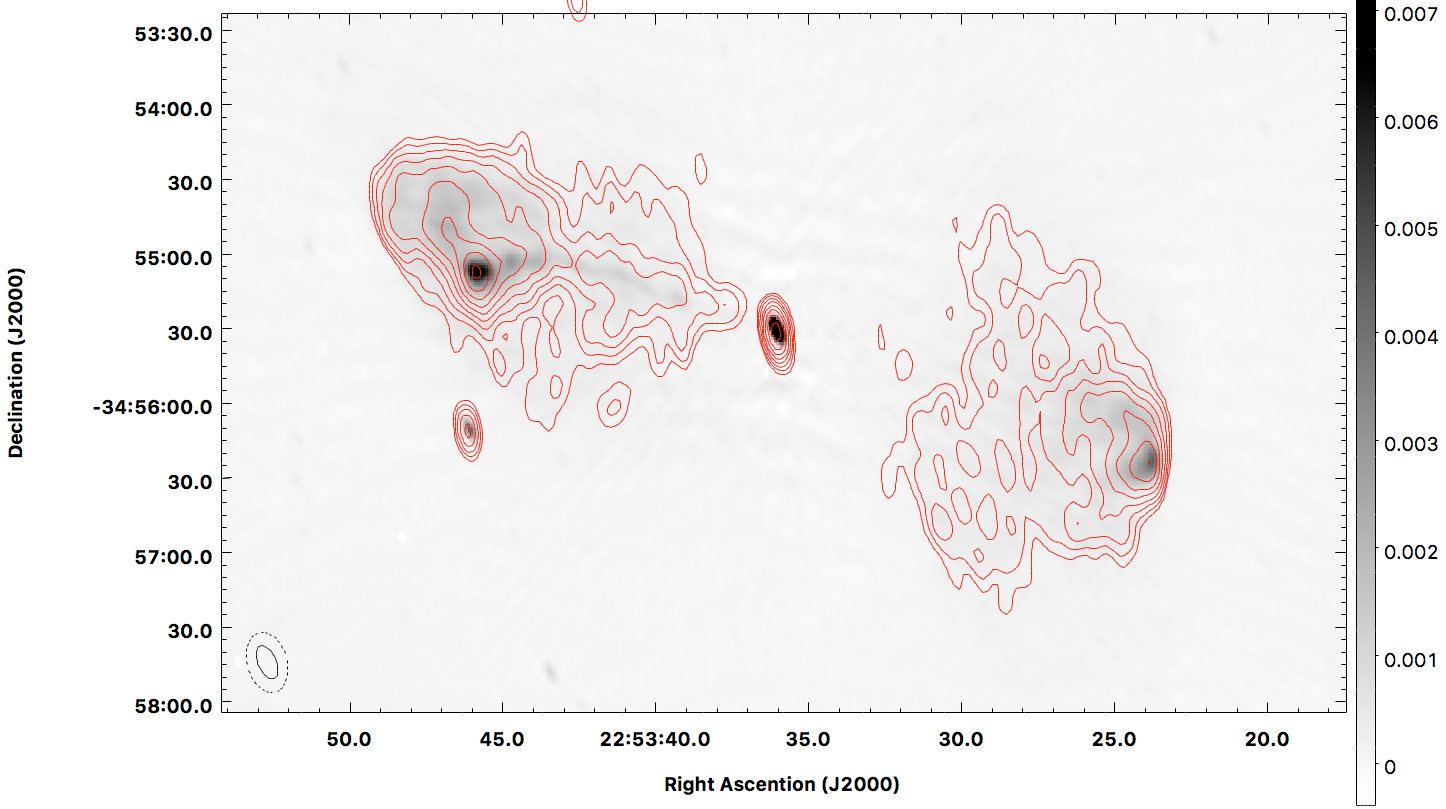}
\caption{Greyscale image of  the band-4 uGMRT image (central frequency 670\,MHz). The greyscale stretch  in  Jy/beam is indicated in the side bar.
The  red contours of the band-3  legacy data image (central frequency 323\,MHz)  and start at 2\,mJy increasing by $\sqrt{2}$.
These images reveals the same structure (the  hotspots, diffuse emission and jet) 
which we see from the ASKAP image in  Figure~\ref{fig:askap} although does not recover the extended emission quite as well. 
 The restoring beams are shown in the lower left for the band-3 (dashed ellipse) and band-4 (solid ellipse) images.}
\label{fig:gmrt}
\end{center}
\end{figure*}

\subsubsection{GLASS Radio Data}

  GLASS is targeting this field over six semesters 
(across 2016-2019) with  
ATCA and will provide images and catalogues of the G23 survey field 
at 5.5 and $9.5\,$GHz. The data were acquired with a 2\,GHz bandwidth at both 
frequencies and with the correlator in 1\,MHz mode. ATCA 
was in a 6\,km and 1.5\,km configuration for $69\%$ and 
$31\%$ of the time respectively. The data used here
has a restoring beam of $4''\times 2''$ (BPA$=0^\circ$), and an RMS 
of $\sim 24\,\mu$Jy/beam at 5.5\,GHz, 
and $3.4''\times 1.7''$ BPA$=0^\circ$ and $\sim 40\,\mu$Jy/beam at 9.5\,GHz. 
The data were processed in the standard fashion (see the Users Guide\footnote{\tt www.atnf.csiro.au/computing/software/miriad/userguide})
with {\tt MIRIAD} \citep{Sault:95} using a method similar to that outlined in \cite{Huynh:15}. The phase calibrator was
PKS 2254$-$367. The bright, compact, 
flat-spectrum core of PKS 2250$-$351 is easily detected in GLASS, but the steep-spectrum diffuse lobes 
are resolved out due to the minimum short baselines of $\sim 100$\,m ($\equiv 3.2\,k\lambda$ at 9.5\,GHz) used in GLASS.
However, the two  hotspots are clearly detected at both frequencies (see 
Figure~\ref{fig:glass}).

We can measure the flux densities of the  hotspots using small ellipses and 
sum up the flux in a similar fashion as with the lobes in the ASKAP images. 
The flux densities are converted to mJy and the uncertainties 
are determined as before,  and presented
in Table~\ref{tab:radio}.

\subsubsection{Additional ATCA Observations}

As the lobes of this radio source were resolved out in the GLASS 
observations we requested ATCA `green time' observations. We 
used the same frequencies and bandwidths as GLASS, but in the compact 
H168 configuration to measure the extended flux densities. These 
observations were taken  on $31^{\rm st}$ November and $1^{\rm st}$ 
December 2018 with about 4 and 6.5\,hours
on source time. The data were reduced in a standard fashion with the 
{\tt MIRIAD} software. 
 PKS B1934$-$638 was used to establish an 
absolute flux density consistent with the \cite{Baars:77} standard
as well as to derive our bandpass correction. 
The data were flagged for  radio frequency interference using the guided automated flagging 
\textsc{pgflag} task.  The bandpass was established using 
the radio spectrum of  PKS B1934$-$638 as a reference. The solutions were then copied over to 
the phase calibrator, PKS 2254$-$367, 
and  a time dependent phase solution was determined. 

Four pointings were used to adequately cover the full extent of 
PKS 2250$-$351 due to the size of the primary beam at 9.5\,GHz. Each 
pointing was imaged independently using the {\tt mfclean} task
to perform image deconvolution while accounting for the spectral variation 
of both the synthesised beam and source intensity across the 2\,GHz bandwidth.
After being primary beam-corrected these four images were 
then mosaiced together. We used a Briggs robustness weighting of 
$R=0$ at $5.5\,$GHz and $R=1$ at $9.5\,$GHz. In the resultant images we obtained an RMS of
$\sim 0.2\,$mJy/beam and $\sim 0.1\,$mJy/beam at $5.5$ 
and $9.5\,$GHz respectively. The restoring beams were $46.8''\times 27.8''$ (BPA$=79.1^\circ$)
and $31.0''\times 18.6''$ (BPA$=73.5^\circ$). 
 We measured the flux densities using the {\tt AEGEAN} package \citep{Hancock:18} and
report these flux densities in Table~\ref{tab:radio}. 
 We conservatively have added in quadrature a $5\%$ absolute flux 
calibration uncertainty \citep[e.g. the ATCA users guide and][]{Partridge:16} 
to the {\tt AEGEAN} uncertainties to account for the ATCA absolute flux calibration.
We note that the core  flux densities measured here with the lower 
resolution configuration agree well with those measured from the higher 
resolution GLASS data.  
The lobe flux densities were derived by subtracting the compact
(i.e. hotspot) emission from the total extended flux densities.

\begin{figure*}[th]
\begin{center}
\includegraphics[angle=0,width=1.3\columnwidth]{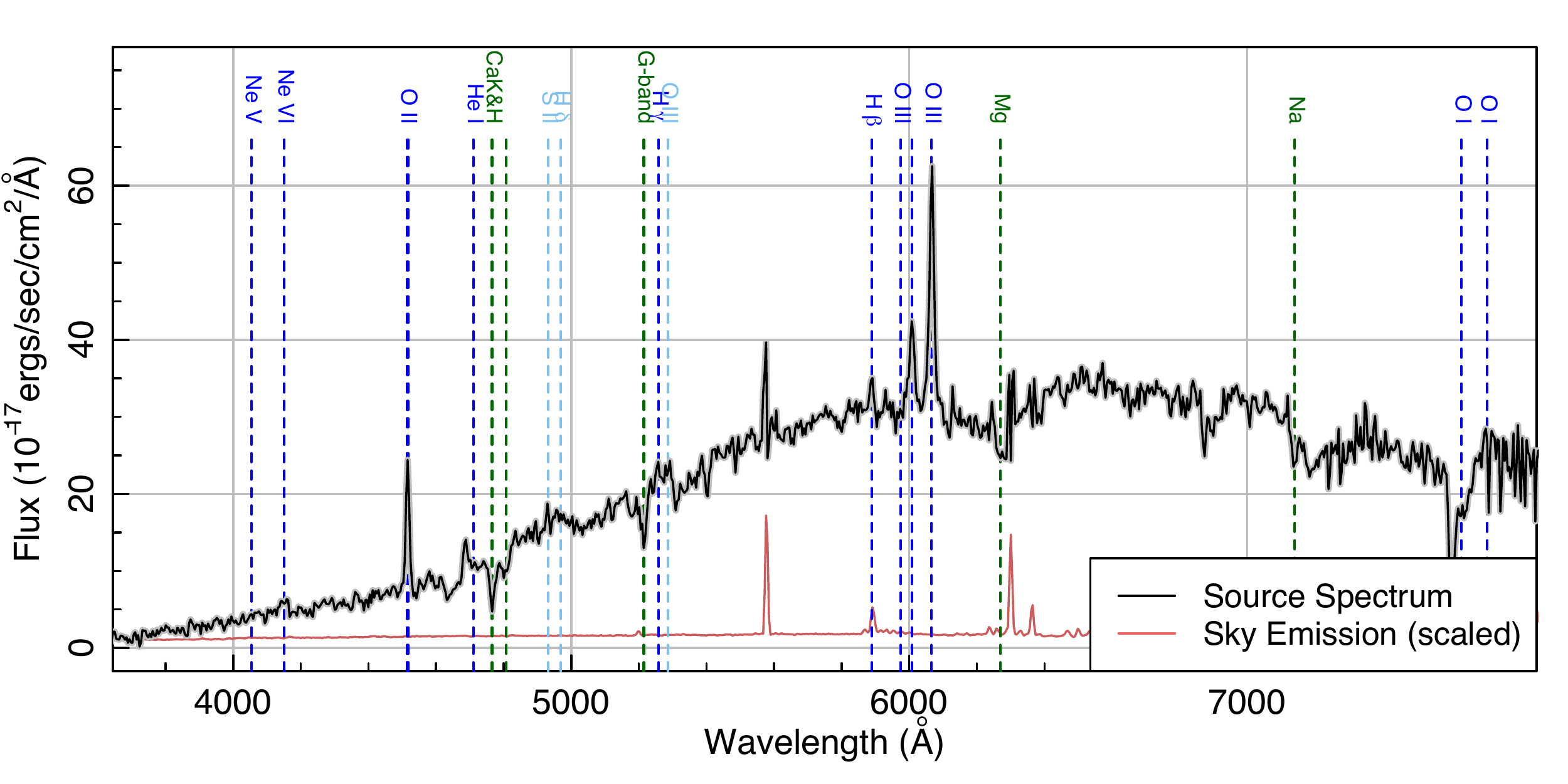}
\includegraphics[trim={0cm 1cm 1cm 0cm},width=0.65\columnwidth,]{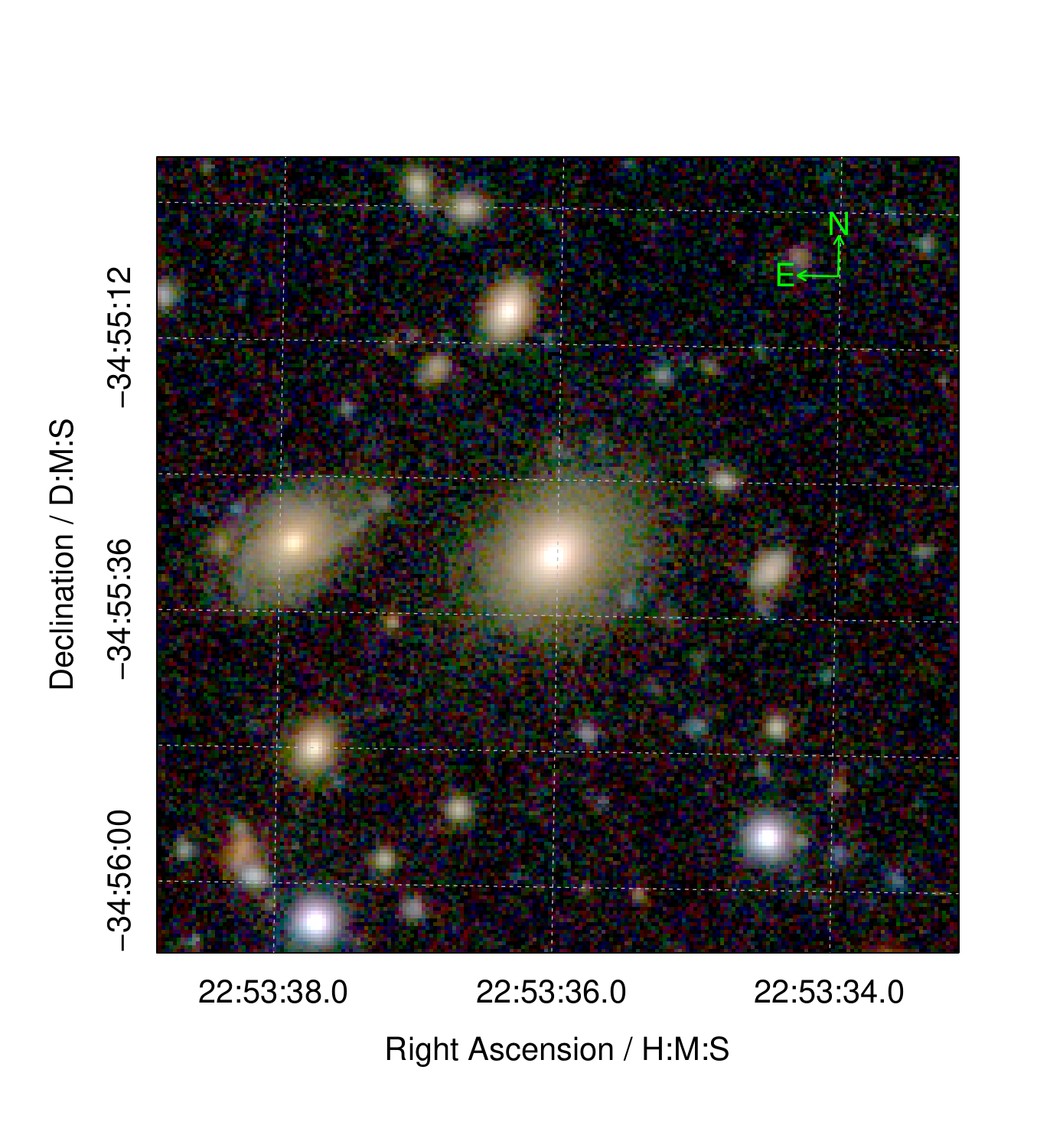}
\caption{(left) The recalibrated 2df spectrum of 2MASS 
J22533602$-$3455305. The vertical dashed lines indicate the observed 
wavelength for spectral lines at $z=0.2115$ and the emission line ratios
are indicative of an AGN (see text for more). (right)
A false colour RGB image of the host galaxy of PKS 2250$-$351 (2MASS 
J22533602$-$3455305). The imaging
comes from GAMA with  $g-$band (blue), $r-$band (green) and $Z-$band (red). 
The host galaxy appears elliptical in shape showing no signs of disturbance. }
\label{fig:host}
\end{center}
\end{figure*}

\subsubsection{uGMRT Radio Data}

Our on-going campaign to study the G23 field with uGMRT includes data in 
both band-3 ($250-500\,$MHz) and band-4 ($550-850\,$MHz). In band-3 the survey consists of 50 
pointings with each pointing observed for about 30 minutes in semi-snapshot mode to 
cover a contiguous $50\,\deg^2$
in total.  The 
band-4 data consists of dedicated pointings of sources of interest
in the G23 field with an ON source time of about 1.5\,hours. The 
data were recorded with the wideband correlator with 200\,MHz 
bandwidth as well as through the narrow band legacy system 
(32\,MHz bandwidth). The band-4
data were processed using a {\tt CASA} based pipeline following procedures 
appropriate for wide band imaging. For band-3, the wide-band data was processed using the  {\tt CASA} based pipeline and the legacy narrow band data was analysed using the {\tt SPAM} pipeline \citep{Intema:09}.

PKS 2250$-$351
is near half-power beam width and since the primary beam correction is not well established for the wide-band at the time of the analysis, we have used the legacy narrow band data at 325\,MHz, instead of 
 the full 
band-3 for flux measurements. The primary beam correction is applied to each pointing before mosaicing the images from the narrow band legacy system. The band-3 image has a best rms of $\sim 100\,\mu$Jy/beam, the legacy 325\,MHz image has rms of  $\sim 300\,\mu$Jy/beam, and the band-4 $\sim 22\,\mu$Jy/beam. The restoring beams are $15.8''\times 6.71''$ (BPA$=9.5^\circ$)  and $7.11''\times 3.59''$ (BPA$=23.7^\circ$)
in band-3 and band-4 respectively  (and the restoring beam of the band-3 legacy image is $12.4''\times 8.0''$, BPA$=15^\circ$). Since PKS 
2250$-$351 is quite bright, it is possible to map the features reasonably well. The uGMRT 
band-4 image is presented in  Figure~\ref{fig:gmrt} with contours from band-3 overlaid. 
 In band-3 the uGMRT is sensitive to scales up to $32'$, 
well beyond the size of this GRG.

To measure the lobe fluxes from the uGMRT we used the same method 
as we did for the ASKAP data employing the same irregular polygons defined by the ASKAP data. The
total flux density and uncertainty were  determined in the same fashion using the restoring beams
given above.
The total  
flux densities of the lobes and core are  presented in Table~\ref{tab:radio}.

\begin{table*}[htbp]
\caption{Radio flux densities of the components of PKS 2250$-$351. 
The GLASS, ATCA green time, ASKAP and uGMRT data 
are presented here for the first time.
The other data here come from AT20G \citep{Murphy:10},  NVSS \citep{Condon:98},
SUMSS \citep{Bock:99}, GLEAM  DR1 \citep{NHW:17}, and  TGSS ADR1 \citep{Intema:17}.
The extended lobe emission seen in the ASKAP and uGMRT images was determined by summing 
the flux density over an irregular polygon measured on the ASKAP image (see text 
for more details). 
For the higher resolution GLASS data we present the flux densities of just the 
 hotspots as the lobes are resolved out. The columns are: radio telescope used, survey the data  are from, the component measured depending on resolution and brightness sensitivity, the 
 observed frequency, the bandwidth, then the flux density (with uncertainty$^a$) of the east lobe, core, west lobe, and the total flux.}
\centering
\begin{tabular}{@{}ccccccccc@{}}
\hline\hline
Telescope & Survey & component & Freq. & $\Delta(\nu)$ & East Lobe & Core & West Lobe & Total \\
          &        &           & (GHz) & (MHz)  & (mJy) & (mJy) & (mJy)    & (mJy) \\
\hline%
\hline%
 ATCA & AT20G & compact & 20 & 512 & $<22^b$     & $57\pm2$ & $<22^b$ & --- \\
\hline%

    & Green Time & extended &    9.5 & 2000 & $22.1\pm 1.2$ & $66.2\pm 3.3$ & $10.7 \pm 0.6$ & $96.6\pm 3.4$\\ 

 ATCA & GLASS & compact        & 9.5 & 2000 & $2.2\pm0.8$ & $61.9\pm 1.3$   & $1.4\pm0.5$ & --- \\
       & --- & lobe$-$compact & 9.5 & 2000 & $19.9\pm 1.4$ & ---           & $9.3\pm 0.8$ & ---\\
\hline%
 ATCA & AT20G$^c$ & compact  & 8.0    & 128 &    ---         & $88\pm6$ & ---  & $88\pm6$\\
 \hline%
 & Green Time & extended  & 5.5 & 2000 & $38.7\pm 2.1$ & $71.4\pm 3.6$ & $23.6\pm 1.4$ & $134\pm 49$\\ 
 ATCA & GLASS & compact  & 5.5  & 2000 & $4.4\pm1.5$   & $66.3\pm 5.6$ & $3.5\pm1.2$ & ---\\
       & --- & lobe$-$compact  & 5.5 & 2000 & $34.3\pm 2.6$ & ----          & $20.1\pm 1.8$ & ---\\
\hline%
ATCA & AT20G$^c$ & compact   & 5.0 & 128  &    ---         & $60\pm4$ & ---  & $60\pm4$ \\ 
\hline%
VLA & NVSS & total   & 1.4  & 42 & $135\pm5$  & $50.0\pm2.2$ & $100.1\pm3.8$  & $285\pm29$\\ 
\hline
ASKAP & EMU  & total & 0.888 & 288 & $193\pm26$ &    n/a$^d$ &   $145\pm17$ &  $>338$\\
\hline%
 MOST & SUMSS & total  & 0.843 & 3 & $175.1\pm5.8$ & $64\pm5.6$ & $153.8\pm7.5$ & $393\pm41$\\ 
\hline%
 uGMRT & GLASS & total & 0.675 & 200  & $237\pm 17$ & $45\pm3$ & $182\pm12$ & $464\pm29$\\
       & GLASS & total & 0.323 & 200 & $523\pm 68$ & $28\pm3$ & $438\pm48$ & $989\pm91$\\
\hline%
  & & & 0.200$^e$& 61.44& $767\pm 62$ & --- & $644\pm 53$ & $1410\pm 82$\\
  & & & 0.227 & 7.68 & $ 671 \pm 57 $ & --- & $ 562 \pm 50 $ & $ 1232 \pm 76 $ \\
 & & & 0.220 & 7.68 & $ 690 \pm 58 $ & --- & $ 578 \pm 51 $ & $ 1268 \pm 77 $ \\
 & & & 0.212 & 7.68 & $ 730 \pm 61 $ & --- & $ 592 \pm 52 $ & $ 1322 \pm 80 $ \\  
 & & & 0.204 & 7.68 & $ 743 \pm 62 $ & --- & $ 639 \pm 55 $ & $ 1382 \pm 83 $ \\  
 & & & 0.197 & 7.68 & $ 755 \pm 62 $ & --- & $ 613 \pm 52 $ & $ 1368 \pm 81 $ \\  
 & & & 0.189 & 7.68 & $ 770 \pm 64 $ & --- & $ 661 \pm 56 $ & $ 1432 \pm 85 $ \\  
 & & & 0.181 & 7.68 & $ 791 \pm 66 $ & --- & $ 670 \pm 57 $ & $ 1461 \pm 87 $ \\  
 & & & 0.174 & 7.68 & $ 844 \pm 70 $ & --- & $ 735 \pm 62 $ & $ 1579 \pm 93 $ \\  
MWA & GLEAM DR1 & total & 0.166 & 7.68 & $ 873 \pm 72 $ & --- & $ 711 \pm 60 $ & $ 1584 \pm 93 $ \\  
 & & & 0.158 & 7.68 & $ 956 \pm 78 $ & --- & $ 751 \pm 63 $ & $ 1708 \pm 100 $ \\  
 & & & 0.151 & 7.68 & $ 979 \pm 80 $ & --- & $ 800 \pm 67 $ & $ 1779 \pm 104 $ \\  
 & & & 0.143 & 7.68 & $ 1038 \pm 85 $ & --- & $ 889 \pm 74 $ & $ 1927 \pm 113 $ \\  
 & & & 0.130 & 7.68 & $ 1069 \pm 90 $ & --- & $ 914 \pm 79 $ & $ 1983 \pm 120 $ \\  
 & & & 0.122 & 7.68 & $ 1145 \pm 97 $ & --- & $ 974 \pm 85 $ & $ 2119 \pm 129 $ \\  
 & & & 0.115 & 7.68 & $ 1264 \pm 107 $ & --- & $ 1025 \pm 89 $ & $ 2289 \pm 139 $ \\  
 & & & 0.107 & 7.68 & $ 1378 \pm 118 $ & --- & $ 1144 \pm 101 $ & $ 2522 \pm 155 $ \\  
 & & & 0.099 & 7.68 & $ 1483 \pm 127 $ & --- & $ 1199 \pm 106 $ & $ 2682 \pm 165 $ \\  
 & & & 0.092 & 7.68 & $ 1531 \pm 131 $ & --- & $ 1301 \pm 114 $ & $ 2832 \pm 174 $ \\  
 & & & 0.084 & 7.68 & $ 1725 \pm 147 $ & --- & $ 1436 \pm 125 $ & $ 3162 \pm 193 $ \\  
 & & & 0.076 & 7.68 & $ 1870 \pm 165 $ & --- & $ 1550 \pm 143 $ & $ 3420 \pm 218 $ \\  

\hline%
 GMRT & TGSS ADR1 & total & 0.1475 & 16.7 & $721\pm72$ & $<20^f$ & $629\pm63$ & $<1379\pm166$\\ 
\hline\hline
\end{tabular}
\label{tab:radio}
\medskip
\tabnote{$^a$ The MWA uncertainties include an additional $8\%$ to account for the overall flux calibration.}
\tabnote{$^b$ $2\sigma$ upper limits from $20\,$GHz image provided by P. Hancock (private communication).}
\tabnote{$^c$ Note AT20G images at 5 and 8\,GHz are not available so there are no constraints on the lobes
at these frequencies}
\tabnote{$^d$ It is not possible to determine the core flux density in the ASKAP image due to bandwidth 
smearing.}
\tabnote{$^e$ 0.2\,GHz MWA flux determined from wide-band ($170-230\,$MHz)
image. The other MWA photometry uses the position in the wide-band image as a
prior  (see \S\ref{sec:radlit}).}
\tabnote{$^f$ Estimated upper limit from visual inspection of image (see \S\ref{sec:radlit}).}
\end{table*}

\subsection{Ancillary Data}

\subsubsection{Literature}
\label{sec:anclit}
 From the bright compact radio emission at the core, we find that
the host galaxy of PKS 2250$-$351 (see Figures~\ref{fig:askap} 
and~\ref{fig:gmrt})  is 2MASS J22533602$-$3455305 \citep{Skrutskie:06}. 
This galaxy has a redshift of $z=0.2115\pm0.0003$ as determined by spectroscopy
obtained as 
part of the 2df Galaxy Redshift Survey \citep[2dfGRS][]{Colless:01}. 
We recalibrated the 2dFRG spectrum so that the $r-band$ magnitude 
derived from it matches the observed GAMA value, including a 
correction  factor of 3.5 for the $2''$ fibre not encompassing the 
entire extent of the galaxy (Figure~\ref{fig:host}, right).
Its spectrum (see Figure~\ref{fig:host}, left panel) 
features prominent narrow [OII] and [OIII] emission lines, but the 
H$\alpha$ line is  redshifted out of the observed wavelength range.

The observed [OIII]$\lambda$5007/H$\beta$ line ratio 
is $\sim 0.9$ which is a relatively high 
value, putting it in the AGN region of [NII]$\lambda$6584/H$\alpha$ v 
[OIII]$\lambda$5007/H$\beta$ parameter space \citep[after][]{Baldwin:81}
for almost all values of [NII]$\lambda$6584/H$\alpha$. 
The bright [OII] and [OIII] lines are narrow ($<1000\,$km/s) which is consistent 
with this galaxy being an obscured AGN. 
This spectrum was obtained 
from a $2''$ fibre, smaller than the total extent of the galaxy ($7-8''$), hence it 
is perhaps not surprising that the nuclear emission lines dominate this spectrum.

\subsubsection{GAMA}

The GAMA survey provides $ugriz$ optical photometry from the VLT Survey Telescope  (VST) and $YJHK$ 
near-IR photometry from the Visible and Infrared Survey Telescope for Astronomy (VISTA)  for the G23 field.
We present the photometric data of PKS 2250$-$351 in 
Table~\ref{tab:gama} measured by the {\tt ProFound} software \citep{Robotham:18}.
In a $zJK$ RGB image (Figure~\ref{fig:host}) the host galaxy
appears extended and elliptical. 
The upcoming public release of these G23 data will 
include a cross-matched catalogue of $\sim$45k
high-quality optical spectra.

\begin{table}[t]
\caption{The GAMA UV to far-IR photometry of 2MASS J22533602$-$3455305, the host galaxy of PKS 2250$-$351.
The {\it WISE} flux densities are converted from Vega magnitudes  as explained in the text. PACS and SPIRE ($100-500\,\mu$m) upper limits are determined from the
public H-ATLAS images at $2.5\times$ the local RMS. The columns are: band, effective wavelength
and observed flux density with uncertainty.}
\centering
\begin{tabular}{@{}lcc@{}}
\hline\hline
Band & effective wavelength & Flux Density \\
    &  ($\mu$m) & ($\mu$Jy) \\
\hline%
 $u_{\rm VST}$  & 0.3581 & $21.6\pm 1.9$ \\
 $g_{\rm VST}$  & 0.4760 & $103\pm 9$ \\ 
 $r_{\rm VST}$  & 0.6325 & $380 \pm 34$ \\ 
 $i_{\rm VST}$  & 0.7599 & $589 \pm 52$ \\ 
 $Z_{\rm VST}$  & 0.8908 & $743 \pm 66$ \\ 
 $Y_{\rm VISTA}$  & 1.023 & $951 \pm 84$ \\ 
 $J_{\rm VISTA}$  & 1.256 & $1,184 \pm 104$ \\ 
 $H_{\rm VISTA}$  & 1.650 & $1,393 \pm 123$ \\
 $K_{\rm VISTA}$  & 2.157 & $1,702 \pm 150$ \\ 
 $W1$ & 3.400 & $868 \pm 49$\\
 $W2$ & 4.652 & $832 \pm 63$\\
 $W3$ & 12.81 & $1261 \pm 424$\\
 $W4$ & 22.38 & $2050 \pm 890$\\
 $P100$ & 98.89 & <5,250\\
 $P160$ & 156.1 & <5,500\\
 $S250$ & 249.4 & <17,000\\
 $S350$ & 349.9 & <13,750\\
 $S500$ & 504.1 & <13,500\\
\hline\hline
\end{tabular}\\
\label{tab:gama}
\end{table}

\subsubsection{{\it WISE} and {\it Herschel} Infra-Red Data}

The {\it Wide-field Infrared Survey Explorer} mission \citep[{\it WISE},][]{Wright:10} imaged 
nearly the entire sky at 3.4, 4.6, 12 and 22\,$\mu$m.
We use the aperture magnitudes of 2MASS J22533602$-$3455305 from the AllWISE 
catalogue\footnote{\tt http://wise2.ipac.caltech.edu/docs/release/allwise/} which 
are more appropriate for a slightly extended source. 
We then converted these to flux densities presented in Table~\ref{tab:gama} 
using the conversion factors from \cite{Jarrett:11} and \cite{Brown:14}.

This photometry can be used to derive the 
following colours: $W1-W2=0.60\pm 0.04$ and $W2-W3 = 2.37\pm 0.32$ (where $Wx$ is the Vega 
magnitude in band $x$,  and 1-3 corresponds to the 3.4, 4.6 and 12\,$\mu$m bands 
respectively). These colours place the host galaxy in the region of parameter space typically
occupied by AGN \citep[due to the hot dust of the torus heated by the 
accretion disc,][]{Jarrett:11,Mateos:12}.

The GAMA survey fields were also covered with far-infrared imaging from the {\it Herschel} 
Astrophysical Terahertz Large Area Survey \citep[H-ATLAS, ][]{Eales:10}. H-ATLAS imaged large
areas of the sky using {\it Herschel's} two imaging instruments (SPIRE and 
PACS\footnote{The Spectral and Photometric Imaging REceiver (SPIRE) is described in 
\cite{Griffin:10} and Photodetector Array Camera and Spectrometer (PACS) in \cite{Poglitsch:10}}). 
The observations were obtained in PACS-SPIRE parallel mode in which both instruments
are used to image the sky simultaneously. The H-ATLAS south galactic pole observations  included the G23 field. Like all the other H-ATLAS observations, data were taken at 100 and 160\,$\mu$m 
with PACS and 250, 350 and 500\,$\mu$m with SPIRE. This source was not detected in any {\it Herschel} band 
(confirmed by visual inspection) and we list the 
$2.5\sigma$ upper limits in Table~\ref{tab:gama}.

\section{Interpretation and Modelling}
\label{sec:modelling}

\subsection{Radio Morphology}

The existence of bright  hotspots in the 5.5 and 9.5\,GHz seen in the 
GLASS data (see Figure~\ref{fig:glass})
suggest PKS 2250$-$351 is a Fanaroff \& Riley (FR) class II radio galaxy \citep{Fanaroff:74},  compatible with its total 
radio luminosity (see \S~\ref{sec:radcomp})
although it is unusual in that we observe some asymmetry in the hotspots. 
Its projected size of 1.17\,Mpc classes it as a giant radio galaxy. 

The ASKAP and uGMRT images reveal more detail on the radio galaxy 
morphology 
confirming the asymmetric  hotspots seen at 5.5 and 9.5\,GHz. The bright  hotspots 
on either side are embedded in diffuse emission which is resolved out in GLASS. 
On the east lobe the diffuse emission extends well past the bright  hotspot whereas
on the western lobe the  hotspot is at the end. 
 We ascribe the asymmetry seen in the hotspots, jet and width of the lobes
to the environment of this giant radio galaxy (see Section~\ref{sec:env}).
We suggest that the prominence of the eastern jet is not due to beaming 
as it would imply a jet direction within $10^\circ$ of the line of sight, and
 thus an intrinsic size $>6.7\,$Mpc.

\begin{figure}[t]
\begin{center}
\includegraphics[trim={5cm 1cm 7cm 3cm},width=0.95\columnwidth,angle=0,]{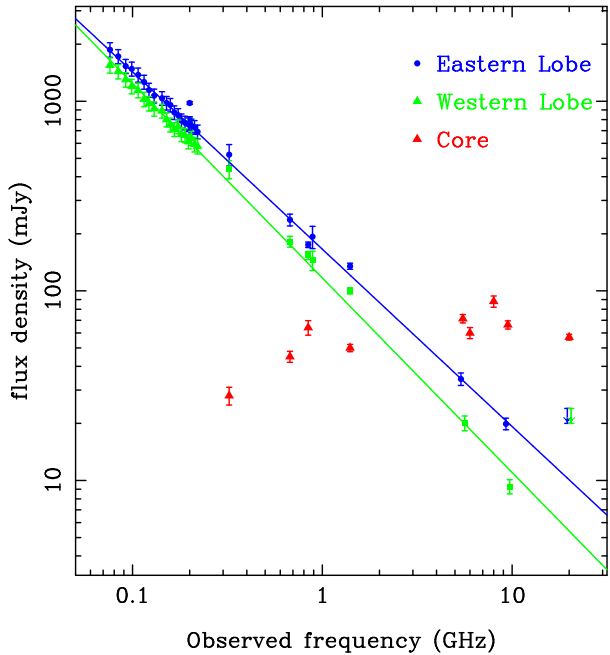}
\caption{The radio (70\,MHz to 20\,GHz) SEDs of the core and each lobe 
of PKS 2250$-$351 as indicated in the insert. 
The lobe SEDs (with the contribution of the  hotspots subtracted) 
are well parameterised as a  single power-law 
 with the best fits overlaid:
$\alpha_{\rm east}=-0.94\pm 0.01 $ 
and $\alpha_{\rm west}=-1.03\pm 0.01$.}
\label{fig:radsed}
\end{center}
\end{figure}

\subsection{Radio Modelling}
\label{sec:radmod}

\subsubsection{Radio component SEDs}
\label{sec:radcomp}

The broad band radio SEDs of the east and west lobes as, well as the core, are presented in 
Figure~\ref{fig:radsed} using the data in Table~\ref{tab:radio}. 
We fit both lobes with a single power-law and a broken 
power-law model\footnote{The break frequency between the power-law slopes was left as a free parameter} (using a least squares method) 
and find that the single power-law is significantly 
preferred \citep[using the Akaike information criterion, corrected for small sample sizes,][]{akaike_new_1974,burnham_model_2002}.  The two 
power-laws are $\alpha_{\rm east}=-0.94\pm 0.01 $  and $\alpha_{\rm 
west}=-1.03\pm 0.01$ and are plotted in Figure~\ref{fig:radsed}.
Note 
that the lobe fluxes have the contributions of the compact emission
(i.e. the hotspots) removed at 5.5 and 9.5\,GHz. This correction
has only a negligible effect on the spectra 
 ($\Delta\alpha\le 0.03$). The slopes of the lobe SEDs are 
steep as expected
from synchrotron emission originating from an aged population of
relativistic electrons with a steep power-law energy distribution.

The uncertainties on these spectral indices are probably 
underestimated due to some correlation between the MWA flux densities
\citep[see][for a full discussion]{Callingham:17}. By eye one 
might argue that there is weak evidence for the west lobe becoming
steeper above $1\,$GHz, but it is possible that some extended 
emission is still resolved out even with the lower resolution ATCA 
observations. Hence, we conclude that the spectral indices
of each lobe are consistent within our overall uncertainties of their
value.

The compact lobe components measured from the GLASS data allow 
us to determine the two-point spectral (5.5-9.5\,GHz) indices of the hot 
spots. We find values of $\alpha_{\rm east}^{\rm hs}=-1.27\pm 0.90$ 
and $\alpha_{\rm west}^{\rm hs}=-1.68\pm0.91$. While these values 
are steeper than those for the lobes they are consistent with the lobe 
values within the uncertainties. 
The hotspots are still detected in the lowest frequency,
high resolution radio image at 323\,MHz from the uGMRT (see Figure
~\ref{fig:gmrt}).
However, the resolution is $\sim 4$ times worse 
compared to the 
GLASS images. Hence, it is difficult to 
quantify the contribution of the hotspots to the lobes  at
low frequency. The hotspots could conceivably 
contribute a larger fraction of the total flux at lower frequencies, 
but equally 
we might expect the  hotspots to become
synchrotron self-absorbed (SSA) at some low frequency. 

The spectrum of the unresolved core appears fairly flat above $\sim1\,$GHz
as expected from the  classical assumption of the  
superposition of many 
separate synchrotron components each with different turn-over frequencies due to SSA.

There is some suggestion
of a downturn at low frequencies from the uGMRT data which is 
further suggested by the faint emission seen in TGSS (see \S~\ref{sec:radlit}).
However, we cannot tell with these data if this is due to SSA or 
free-free absorption processes. 
The AT20G $8\,$GHz 
flux density is a little higher than the other measurements, but 
as it was observed near-simultaneously with the $5\,$GHz data we put 
this down to measurement scatter or genuine variability of the core. 

Using the  lobe spectral indices for the k-correction, we find a 
total luminosity at 1.4\,GHz (151\,MHz) of $L_{\rm 1.4GHz}=3.24\pm 0.25\times 
10^{25}\,$W\,Hz$^{-1}$ ($L_{\rm 151MHz}= 2.70\pm 0.11 \times 10^{26}\,$W\,Hz$^{-1}$) consistent 
with  PKS 2250$-$351 being an FR II source as determined from the observations of  hotspots (see  Figure~\ref{fig:askap}). 

\subsubsection{Jet kinetic power and source age}
\label{sec:jetPower}

We use observations of the symmetric western lobe to infer jet kinetic power and source age using the  Radio AGN in Semi-analytic Environments (RAiSE) dynamical model \citep{Turner:15,Shabala:17,Turner:18a,Turner:18b}. 
We refer the interested reader to those papers for a comprehensive description of our modeling approach. Briefly, we produce luminosity-size tracks and optically thin synchrotron radio continuum spectra, for a wide range in jet kinetic power. While in some sources it is possible to infer the lobe magnetic field strength from the break in the synchrotron spectrum (Turner et al. 2018b), 
 we find no break in the SED of PKS 2250$-$351 (Figure~\ref{fig:radsed})
and hence we adopt a value of 0.3 times the equipartition field. This is characteristic of FR-II lobe field strengths measured through inverse-Compton observations \citep{Croston:17}.
Modelled jets consist of a pair-plasma (i.e. no protons), and we use the observed low-frequency spectral index to constrain the initial particle spectral index at the  hotspots, $\alpha_{inj}=-0.6$.
 The injection spectral index is related to the power-law index in energy ($s$) as $\alpha_{inj}$$=$$\frac{-(s-1)}{2}$. Thus, the $\alpha_{inj}$ value we 
use corresponds to $s$$=$$-2.2$ which is consistent with expectations from diffusive shock acceleration \citep[e.g.][]{Heavens:88}. The lobes
have a steeper spectrum than this $\alpha_{inj}$ most likely due to energy losses \citep{Turner:18b}.

Backflow of accelerated plasma from the  hotspots inflates the radio cocoon, which expands supersonically 
through the intracluster gas. RAiSE models the dynamics of lobe expansion (radially and transverse), and accounts for synchrotron, adiabatic and inverse-Compton losses from the emitting electrons. We run a grid of models for 
environments corresponding to group-centre galaxies in haloes  hot intra-cluster medium (ICM) with mass 
spanning $3 \times 10^{13} - 3 \times 10^{14} M_\odot$; we consider these to be representative upper and lower 
limits of the (unknown) ICM conditions in the outskirts of Abell 3936. We then used maximum likelihood to find 
the best-fitting jet kinetic powers and dynamical ages. Regardless of environment, we consistently recover a total 
(i.e. for both jets)  lobe-derived jet power of $Q_{\rm jet}^{\rm lobe}=1-1.5 \times 10^{38}\,$W, and ages of 
$260$-$320\,$Myr. These values do not change appreciably even if we relax our assumptions about the magnetic 
field. 
 This age is quite large compared to other GRGs \citep{Ishwar:99} and even `old' regular radio-loud AGN \citep{Murgia:11}.

Lobe dynamics trace time-averaged jet power, which may not be directly 
comparable with the (quasi-instantaneous) accretion rate determined 
through mid-IR observations. 
\cite{Godfrey:13} presented a method for calculating the 
instantaneous\footnote{Assuming the jet has a velocity of 
$\sim 0.7\,c$, the time it takes to reach the  hotspots is a few Myr, 
much shorter than the age of the lobes.} 
jet power from observations of  hotspots; this method has been shown to 
be consistent with time-averaged dynamical jets powers 
\citep{Shabala:13}
with some scatter, as expected. We obtain a  hotspot-derived jet power of 
$Q_{\rm jet}^{\rm hs}\sim 8 \times 10^{36}$\,W
for each jet, i.e. $1.5 \times 10^{37}$\,W in total, an order of magnitude lower than the time-averaged  lobe-derived jet power. We suggest some plausible explanations for this interesting discrepancy in the discussion (Section~\ref{sec:disjet})  below.

\begin{figure}[t]
\begin{center}
\includegraphics[angle=270,trim={3cm 1cm 1cm 3cm},clip,width=\columnwidth,]{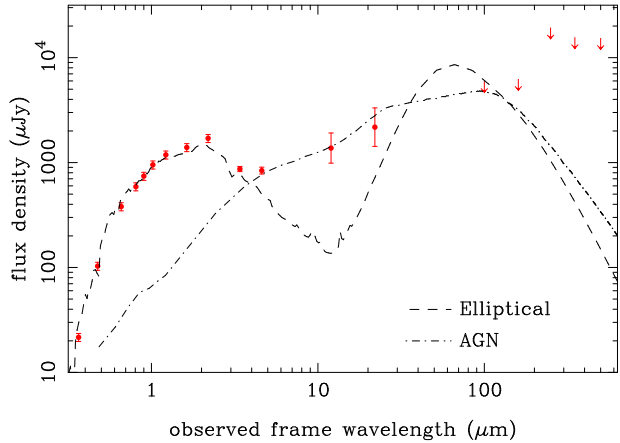}
\caption{We plot the UV to far-IR SED of 2MASS J22533602$-$3455305, the host galaxy of PKS 2250$-$351, 
with the data from Table~\ref{tab:gama}. 
We show the best fit stellar SED model  (dashed line)
to these data
(excluding the three longer wavelength {\it WISE} bands which are dominated by the AGN).
We overlay an AGN model  (dash-dot line) from \cite{Symeonidis:16} 
to demonstrate how it dominates the {\it WISE} bands. }
\label{fig:gama}
\end{center}
\end{figure}

\subsection{The Host Galaxy}

The radio emission from the bright compact core is unambiguously 
associated with the galaxy 2MASS J22533602$-$3455305.
Its morphology and colours indicate it would normally be a `red and dead' massive 
elliptical.  From Figure~\ref{fig:host} it appears the major axis
is close to perpendicular to the radio jet axis, a feature which is
common in elliptical hosts of radio galaxies \citep{Battye:09}.

\begin{figure}[t]
\begin{center}
\includegraphics[width=1.0\columnwidth]{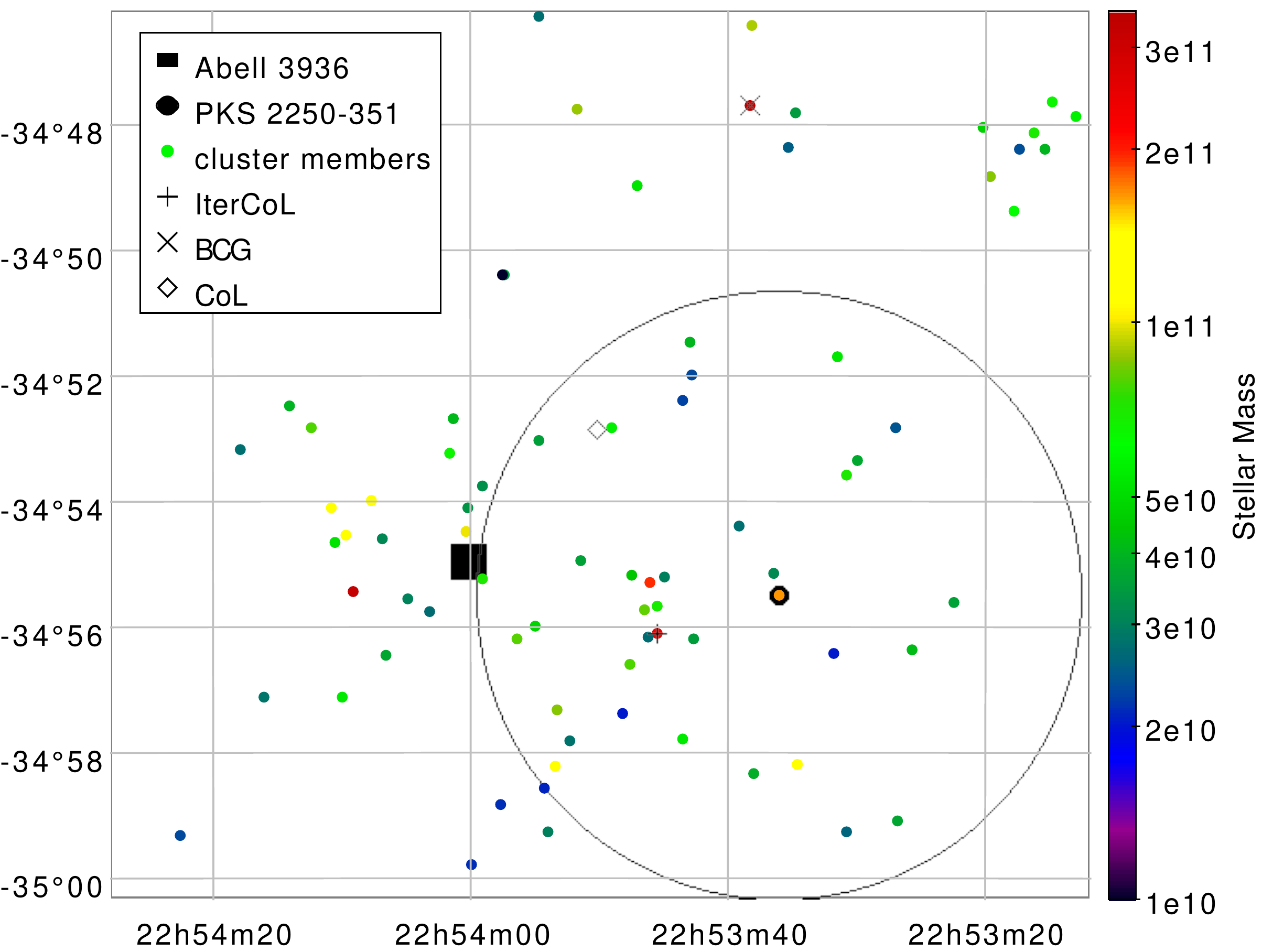}
\caption{
Sky distribution of spectroscopically confirmed GAMA sources lying at
$0.207\le z\le 0.2180$ (i.e. $\Delta(cz)\le 1,500\,$km\,s$^{-1}$). 
PKS 2250$-$351 is indicated by a larger black dot surrounded
by   a 1\,Mpc radius circle. The large black square is the 
Abell cluster position and the `$+$', `X' and diamond indicate
three different estimates of the cluster centre from the GAMA 
group catalogue. The colour code of the galaxies indicate their 
stellar masses. The radio galaxy lies to the west of our most
confident estimate of the cluster centre, the Iterative Centre
of Light (`IterCoL').}
\label{fig:zdist}
\end{center}
\end{figure}

\subsubsection{SED Modelling}
\label{sec:hostmod}
We fit the UV to far-IR SED using the {\tt ProSpect}\footnote{\tt https://github.com/asgr/ProSpect} code. 
The {\tt ProSpect} code fits stellar libraries for different populations and dust 
templates to observed data. The infra-red emission from the dust is balanced by the
absorption fitted to the optical/UV photometry assuming a uniform screen. 
As the three longer wavelength {\it WISE} bands are dominated by the AGN (as indicated by the 
{\it WISE} colours, see Section~\ref{sec:anclit}) we do not use them in the fitting of the stellar components. 
To account for the non-detections in the {\it Herschel} bands we use a range of flux values from $0-2.5\,\sigma$,
i.e. input flux densities of $1.25\pm 1.25\,\sigma$ (where $\sigma$ is the 
image RMS).
The SED is best fit with an old stellar population which had a peak star formation 
rate (SFR) around $5\,$Gyr ago. 

We find the galaxy has a negligible  current SFR of 
$\ll 0.5\,M_\odot\,$yr$^{-1}$ 
and a stellar mass of $1.93\pm 0.07\times 10^{11}\,M_\odot$. This fit is shown in  Figure~\ref{fig:gama}
where the {\it Herschel} bands are shown as $2.5\sigma$ upper limits. We overlay an AGN template from
\cite{Symeonidis:16} to demonstrate the AGN dominance in the {\it WISE\,} bands. 
In comparison to the general population of galaxies 
these values put this galaxy far below the SFR/stellar mass `main sequence' 
\citep{Brinchmann:04,Noeske:07,Seymour:08} with a specific SFR of $<< 1\times 10^{-2}\,$Gyr$^{-1}$  (i.e. `red and dead').

The non-detections in the far-IR from the {\it Herschel Space Observatory} (see table~\ref{tab:gama})
 imply an upper limit to the $60\,\mu$m luminosity of 
$L_{60\mu m}\lessapprox  1\times 10^{10}\,L_\odot$ which corresponds to a SFR 
$\lessapprox 2\,M_\odot$yr$^{-1}$ \citep[using the relationship from][]{Calzetti:10} and are therefore consistent with the low reported SFR. The host galaxy is also undetected in the 
relatively shallow {\it ROSAT} all-sky X-ray 
survey \citep{Voges:99} and 
 has yet to be targetted by other X-ray facilities.

\subsubsection{Black Hole Accretion Rate}

The {\it WISE} colours and luminosity, along with the optical spectrum, 
are all consistent with on-going obscured accretion onto a  SMBH 
and with the lack of star formation indicated by the UV to 
near-infrared SED. 
Hence the {\it WISE} photometry can be used to estimate a mid-infrared 
luminosity and infer an accretion rate. 
Interpolating between the $W2$ and $W3$ flux densities (assuming a power-law) 
we can measure the $5\,\mu$m rest-frame luminosity as 
$\nu L_{\rm 5\mu m}=1.560\pm 0.16 \times 10^{10}\,L_\odot$ in solar luminosities. 
This is around an order of magnitude below the knee
of the low-redshift luminosity function of mid-infrared selected AGN \citep{Lacy:15}. Following the conversion factor of ten from \cite{Lacy:15} 
we find $L^{AGN}_{BOL}=1.56\pm 0.16\times 10^{11}\,$L$_\odot$.

We also estimate the bolometric AGN luminosity from the 
[OIII]$\lambda$5007 line. We  measured an [OIII]$\lambda$5007 flux of 
$1.3\pm 0.3\times 10^{-15}\,$erg\,s$^{-1}$\,cm$^{-2}$ which corresponds to a 
luminosity of $L_{\rm OIII\lambda 5007}=3.4\pm 0.9 \times 10^7\,$L$_\odot$.
Following \cite{Heckman:04} we convert to a bolometric 
AGN luminosity by multiplying by a factor of $\sim 3500$ ($\pm 0.38\,$dex), thus obtaining 
$L^{AGN}_{BOL}=1.2 \pm 0.3 \times 10^{11}\,$L$_\odot$. This value 
is consistent with the mid-IR derived AGN luminosity
and we take the mean of these values ($L^{AGN}_{BOL}\sim 1.38\pm0.18\times 10^{11}\,$L$_\odot$) 
in our deliberations below (where the uncertainty is taken from the range of the 
two values).

The efficiency of black hole accretion in AGN is not strongly 
constrained, but is  estimated to be between 6\% to 40\% 
\citep[see discussion in Section 6 of][]{Drouart:14}. Here we 
conservatively take a value of 10\%, but note it can vary. Using this 
value we obtain an accretion rate of $\sim 0.1\,$M$_\odot$yr$^{-1}$. We 
can compare this value to 
the Eddington accretion rate which depends on black hole 
mass\footnote{$\frac{L_{\rm EDD}}{L_\odot}=3.2 \times 10^4 
\big(\frac{M}{M_\odot}\big)$, where $M$ is the mass of the SMBH.}. 
We  estimate the mass of the black hole from the mass of the host 
galaxy assuming it lies on the local M$-\sigma$ relation \citep[e.g.][]{Ferrarese:00,Kormendy:13}. Specifically we use the stellar mass reported above as the bulge 
mass, as our host appears to be a pure elliptical, and the conversion to black hole mass
quoted in \cite{Haring:04}. We therefore estimate a black hole mass of $M_{BH}= 3.3\pm 0.8\times 
10^8\,M_\odot$ \citep[where the uncertainty comes from that in the conversion equation of][]{Haring:04} which we can use to determine an 
Eddington luminosity of $\sim 1\times 10^{13}\,L_\odot$. Hence our estimated Eddington accretion 
rate is $\lambda_{\rm EDD}\sim \frac{1.4\times 10^{11}}{1\times 10^{13}} \sim 0.014\pm0.004$ 
where the uncertainty is propagated from those of the black hole mass and bolometric AGN luminosity.

\subsection{Environment}
\label{sec:env}

 PKS 2250$-$351 has been associated with the 
`Irregular' cluster Abell 3936 \citep[by][]{Brown:91} 
with an Abell count of 95 and richness of `2'. 
We investigate this association using a galaxy group catalogue 
(included as part of the forthcoming G23 field data release)
based on a friends-of-friends algorithm following the method of
\cite{Robotham:11}. This group 
catalogue was produced with the $\sim 45$k high 
quality spectroscopic redshifts in G23 and
includes information such as the number of members, the virial mass 
and three separate estimates of the group centres. These centres are 
based on a
centre of light (CoL) method, an iterative centre of light method 
(IterCoL) and the brightest cluster galaxy (BCG).

We confirm 2MASS J22533602$-$3455305 as a member of the 
largest group in G23 at the reported position of Abell 3936
(see Figure~\ref{fig:zdist}).
This group has 92 spectroscopic members (including 2MASS 
J22533602$-$3455305) and an estimated virial mass of 
$9.7\pm1.0\times10^{14}\,M_\odot$. This mass estimate is based on the 
observed velocity dispersion ($705\pm 50\,$km\,s$^{-1}$) 
and the estimated cluster radius. Note the 
uncertainty in the mass is from the velocity dispersion and does not 
include that on the cluster radius.
However, the number of members and virial mass estimate strongly 
suggest that this is a massive cluster at $z=0.213$ \citep[a 
higher redshift than estimated by][]{Brown:91}

In Figure~\ref{fig:zdist}, we plot the spatial distribution of the 
spectroscopically confirmed galaxies in the range
$0.207<z<0.218$  (i.e. within $\Delta(cz)=1,500\,$km\,s$^{-1}$ of 
the $cz$ of 2MASS J22533602$-$3455305). For 
reference we indicate a circle of radius 1\,Mpc around PKS 2250$-$351. The colour-bar 
indicates the estimated stellar masses of the galaxies (determined as described in Section~\ref{sec:hostmod}). The three group centre estimates from the group catalogue are 
indicated as shown in the legend. 
The results of these three methods differ significantly. 

The IterCoL estimate lying due east of PKS 2250$-$351 is probably the best if one were to 
make a judgement from the distribution of galaxies in Figure~\ref{fig:zdist}. The CoL 
method is further north, drawn in that direction by the distribution of the galaxies 
in a filament that runs through the whole G23 field. The BCG estimate is clearly 
a long way off for this structure. Hence, we conclude that PKS 2250$-$351 
is a member of Abell 3936, but $\sim$1\,Mpc west of its centre.

\section{Discussion}
\label{sec:dis}

\subsection{The host galaxy}

 For the first time we have identified the radio source PKS 2250$-$351 
with the host galaxy 2MASS J22533602$-$3455305,
a massive elliptical at $z=0.2115$. It has a 
negligible star formation rate but prominent mid-IR emission due to 
accretion onto the  SMBH. The AGN emission is highly obscured in 
the UV and optical, 
consistent with the central engine being obscured by a torus structure 
viewed from  approximately side-on and with the standard 
unification of AGN by orientation \citep{Urry:95}. 

\subsection{Jet Power and Ages}

Our estimate of the age of radio source is relatively well defined. 
PKS 2250$-$351 is an FR-II with well-defined bow-shocks, 
so we know it must still be expanding supersonically 
(and was  expanding faster in the past). For a cluster sound 
speed of  $1000\,$km\,s$^{-1}$  \citep[$\sim 1\,$kpc/Myr,][]{Sarazin:88}
this places  a pretty strong upper limit on the age of 
$\sim(600/M_{\rm lobe})\,$Myr where $M_{\rm lobe}$ is 
the average external Mach number of the lobe-driven 
bow shock (w.r.t. the ICM) over the lifetime of the 
source. 

If the radio 
source is not in the plane of the sky (i.e. is larger)
this would imply a larger age.  The jet power  would then
also need to be adjusted appropriately.
We can estimate the
angle of the jet relative to the line of sight 
using equation 2 of \cite{Hardcastle:98} which relates
the  jet-to-counterpart flux ratio to the  angle between 
the jet and the line of sight. Using the 
ASKAP image and comparing the measured flux in  parts 
of the jet away from knots with an equidistant 
position on the opposite (western) side we 
 obtain a brightness ratio of 3. Assuming
a jet spectral index of $\alpha_{\rm jet}=-0.5$ 
 (from the canonical $s=-2$ electron energy power-law 
index from first-order Fermi acceleration processes)
and a jet speed of $\beta=0.7$\footnote{This is a lower
limit considering that the FRII jets must be substantially 
supersonic and the sound speed is $\sim\frac{1}{3}c$.}, 
this flux density 
ratio is equivalent to $74^\circ$ to the line of 
sight (i.e. close to side on).

Our assumptions also depend on magnetic field 
strength. We took the typical particle to 
magnetic-field energy density ratio from inverse-Compton observations of \cite{Hardcastle:10}.
Lower magnetic field strengths require a higher jet 
power to obtain the same radio luminosity - and hence 
younger age, to match the size constraint. Changing 
the magnetic field by a factor of three (i.e. the 
magnetic field energy density by  a factor of 10, now 
consistent with lobes at equipartition) changes the 
best-fit jet power by 0.2\,dex (giving a broader 
range of $1-2.5\times 10^{38}\,$W), and increasing the 
age range by  30\,Myr (giving a range of $230$-$320\,$Myr).

\subsection{Jet Power vs Accretion Rate}
\label{sec:disjet}

Our results in Section~\ref{sec:jetPower} present a possibly 
conflicting scenario with 
different jet power  estimates. The lobe-derived estimate
of the jet power is an order of magnitude greater than the hotspot-derived value.
This difference is right at the limit of the scatter 
seen between instantaneous jet power estimates from  hotspots and the 
time-averaged dynamic jet powers \citep{Shabala:13}. However,
 that scatter applies to 
younger and smaller radio galaxies where one might expect 
the instantaneous and time-averaged values to 
agree more closely. On the assumption that the 
 change in jet power we observe 
is real, we consider other possibilities. 

Firstly, the AGN accretion rate (and therefore jet power) is simply varying over the age of the 
radio galaxy. While simulations by \cite{Novak:11} have demonstrated rapid, $\ll 1\,$Myr,
variability at low Eddington accretion rates, simulations
by \cite{Gaspari:13} have  shown that `chaotic cold accretion' (CCA)
can lead to variable accretion rates.
In this picture, condensation of hot intracluster gas and subsequent inelastic collisions of cold gas 
clouds remove angular momentum, raining cold clouds onto the super-massive black hole at rates 
significantly in excess of the Bondi rate; this generates 
the powerful jets  averaged over this stochasticity. Mechanical 
feedback from the jets then truncates the cooling process, and the jets are powered by the (less 
efficient) standard cold, thin disc (TD). A testable prediction of this CCA hypothesis is the existence of 
old filamentary gas structures in the vicinity of the AGN host.

Another possibility for the drop in jet power over the life time of the radio source could be 
a change in accretion state. Our estimated Eddington accretion rate of 
$\lambda_{\rm EDD}=0.014\pm 0.004$ is around the value at which accretion onto black holes
switches from an inefficient,  advection dominated thick-disc accretion flow \citep[ADAF,][]{Fabian:95}  
to an efficient  TD accretion flow. An ADAF is more efficient than a TD
at producing jets \citep[e.g.][]{Meier:02}, hence the higher average power derived from the lobes could 
have occurred in this ADAF mode. Then as the accretion rate gradually increases, the state of the accretion disc 
switches to the efficient thin-disc mode and the power of the jet drops by an order of magnitude. 
This is consistent with the relationships between jet power and accretion rate presented in 
\cite{Meier:02} which give a jet power around an order of magnitude 
lower for a TD compared to an ADAF (for the same mass, accretion
rate, and spin).

\subsection{Environment}

 PKS 2250$-$351 resides in what is likely a massive cluster, Abell 3936
(with an estimated virial halo mass of $9.7\pm1\times 10^{14}\,M_\odot$ 
from its velocity dispersion).
It is not uncommon to find GRG in such 
environments \citep{Komberg:09}.
However, to exist at all in such an environment 
this radio galaxy must have a powerful jet and be long-lived, as is 
supported by our modelling of the radio lobes in 
Section~\ref{sec:radmod}. 

 Abell 3936 is not detected in cluster surveys with {\it ROSAT} \citep[e.g.][]{Bohringer:13} 
to a level of $1.8\times 10^{-12}\,$erg\,s$^{-1}$\,cm$^{-2}$ from 0.1 to 2.4\,keV. At the redshift of Abell 3936 this flux is equivalent to a
luminosity of $2.34\times 10^{44}\,$erg\,s$^{-1}$. Using equation~10 of 
\cite{Bohringer:13}\footnote{This relationship assumes no evolution
in the relationship between X-ray luminosity and cluster temperature.} 
the {\it ROSAT} non-detection implies a mass 
$\le 5.5\times 10^{14}\,M_\odot$ (with a $30\%$ uncertainty).
This discrepancy between the `virial' mass and the X-ray mass upper 
limit could be explained if the cluster was not yet relaxed.

This assertion is supported by the observation that the group catalogue 
does not provide  a clear determination of  a cluster centre with 
different estimates providing different positions (see 
 Figure~\ref{fig:zdist}). 
From our analysis in Section~\ref{sec:env} it seems that PKS 
2250$-$351 is probably to the west of the densest 
grouping of galaxies
eastern lobe pointing roughly towards it. 
While the ICM in this case maybe not be as dense as for
a virialised cluster it likely increases towards this 
galaxy concentration. 
Hence, the western radio lobe expands outward more easily as the density of the ICM decreases,
while the eastern lobe is more confined by 
an increasingly dense ICM. 
This idea is further supported by the difference in the lobe
flux densities. A lobe expanding into a less
dense environment will have a lower luminosity for the same jet power, which is consistent with the
eastern lobe being $\sim 30-100\%$ brighter than the western lobe.

The prominence of the jet in the eastern lobe is curious. It could partly be explained by the increasingly 
dense ICM into which it is drilling. 
The knots and kinks seen in the eastern jet could possibly be due to hydrodynamical 
instabilities.
We note that it is  unlikely to be due to relativistic 
beaming which would require a  small angle, $\ll 10^\circ$, between the jet motion and the 
line of sight, inconsistent with our earlier estimate of $74^\circ$. 
We can also rule out precession due to the asymmetry of the lobe.

The jet and lobe asymmetry could be  due to a more complex
morphology  (e.g. a wide-angled tail radio source) 
viewed in projection, but this is also  in contradiction with our 
determination that this radio galaxy is close to side-on.
Future polarimetric observations with ASKAP may better constrain
the orientation of PKS 2250$-$351.
If one radio lobe is expanding into a denser part 
of the ICM one might expect it to have a higher rotation measure due 
to the denser foreground plasma. 

Future X-ray observations (e.g. by {\it eROSITA}) will provide  additional 
clues on these issues, potentially providing evidence of the 
interaction of the radio lobes with the ICM.

\section{Conclusions}
\label{sec:con}

We have completed a detailed study of the giant radio galaxy PKS 2250$-$351, its host
galaxy and environment. 
This work demonstrates how we can improve our understanding of 
 radio-loud AGN using 
broad-band radio surveys (i.e.  from MWA, uGMRT, ASKAP 
and ATCA) with high-quality multi-wavelength surveys such as GAMA. 
In future papers this work can be extended to much larger samples of radio sources  in the G23 field.
By comparing the instantaneous and past averaged jet powers with 
the current accretion rate we can begin to determine when and how jets 
form in super-massive black holes. 

The primary results from this work are: 

\begin{itemize}
\item  We have confirmed the association of the giant radio 
galaxy PKS 2250$-$351 at $z=0.2115$ with the large, but 
irregular Abell 3936 cluster. This cluster is likely unrelaxed and has
no clear centre, although PKS 2250$-$351 lies to the west of the
highest concentration of galaxies.

\item Its host galaxy, 2MASS J22533602$-$3455305, 
 confirmed here for the first time, 
is a massive, `red and dead' elliptical well below the 
SFR/stellar mass main sequence.

\item The OIII emission line and {\it WISE} photometry
imply a current AGN bolometric 
luminosity of $1.38\pm 0.18\times 10^{11}\,L_\odot$ ($\equiv 0.1\,M_\odot$yr$^{-1}$ accretion rate). This AGN 
activity is highly obscured in the UV and optical. We estimate a 
black hole mass of $3.3\pm 0.8\times 10^{8}\,M_\odot$ (from the local $M-\sigma$ relation) 
and hence an Eddington accretion rate of $\lambda_{\rm EDD} = 0.014\pm 0.004$, 
a value close to the transition between an accretion disc in an ADAF state and a thin disc state. 

\item  The lobe-derived  jet power is an order of
magnitude greater than the hotspot-derived jet power.
Given the  hotspots trace the much more recent accretion 
 (a few Myr) and the age of the radio emission is quite old
($260$-$320\,$Myr) we suggest that the accretion disc may have changed 
state over this  period.
We propose that initially this radio galaxy was in 
an inefficient, ADAF mode which 
explains the high, average jet power estimated from the lobes.
With  an increase in accretion rate, the radio galaxy then switched to a thin-disc mode, which 
explains the lower, current jet power estimated from the  hotspots, but also the 
high  current accretion rate seen in the host. 

\item The asymmetry of the lobe widths may possibly be due to a density 
gradient in the ICM  with the less luminous lobe expanding more 
widely into a less dense environment. 
However, without  deeper X-ray observations it is not 
possible to accurately determine the location and position of the centre of the cluster Abell 3936 relative to PKS 2250$-$351.

\end{itemize}

\begin{acknowledgements}
 We thank the referee for the many useful comments provided to improve this paper. 
We thank Martin Krause and Hans B\"ohringer for useful discussions. 
We thank Paul Hancock for a copy of the unpublished AT20G 20\,GHz image of this source. 
We also thank  Jamie Stevens, for assistance
with the ATCA `green time' observations. 
 SS thanks the Australian Government for an Endeavour Fellowship, during which part of this work was completed.
H.A. benefited from
grant CIIC 218/2019 of Universidad de Guanajuato. 
Partial support for L.R. comes from U.S. National Science Foundation Grant AST 17-14205 to the University of Minnesota. The GAMA Legacy ATCA Southern Survey (GLASS) was conducted with the Australia Telescope Compact Array,
which is part of the Australia Telescope National Facility which is funded by the Australian Government for operation as a National Facility managed by CSIRO.

The Australian SKA Pathfinder is part of the Australia Telescope National Facility which is managed by CSIRO. Operation of ASKAP is funded by the Australian Government with support from the National Collaborative Research Infrastructure Strategy. ASKAP uses the resources of the Pawsey Supercomputing Centre. Establishment of ASKAP, the Murchison Radio-astronomy Observatory and the Pawsey Supercomputing Centre are initiatives of the Australian Government, with support from the Government of Western Australia and the Science and Industry Endowment Fund. We acknowledge the Wajarri Yamatji people as the traditional owners of the Observatory site.

This scientific work makes use of the Murchison Radio-astronomy Observatory, operated by CSIRO. We acknowledge the Wajarri Yamatji people as the traditional owners of the Observatory site. Support for the operation of the MWA is provided by the Australian Government (NCRIS), under a contract to Curtin University administered by Astronomy Australia Limited. We acknowledge the Pawsey Supercomputing Centre which is supported by the Western Australian and Australian Governments.

We thank the staff of the GMRT that made these observations possible. GMRT is run by the National Centre for Radio Astrophysics of the Tata Institute of Fundamental Research.

GAMA is a joint European-Australasian project based around a spectroscopic campaign using the Anglo-Australian Telescope. The GAMA input catalogue is based on data taken from the Sloan Digital Sky Survey and the UKIRT Infrared Deep Sky Survey. Complementary imaging of the GAMA regions is being obtained by a number of independent survey programmes including GALEX MIS, VST KiDS, VISTA VIKING, WISE, Herschel-ATLAS, GMRT and ASKAP providing UV to radio coverage. GAMA is funded by the STFC (UK), the ARC (Australia), the AAO, and the participating institutions. The GAMA website is http://www.gama-survey.org/. Based on observations made with ESO Telescopes at the La Silla Paranal Observatory under programme ID 179.A-2004. Based on observations made with ESO Telescopes at the La Silla Paranal Observatory under programme ID 177.A-3016. 

The {\it Herschel}-ATLAS is a project with {\it Herschel}, which is an ESA space observatory with science instruments provided by European-led Principal Investigator consortia and with important participation from NASA. The H-ATLAS website is {\tt http://www.h-atlas.org/}.This publication makes use of data products from the {\it Wide-field Infrared Survey Explorer}, which is a joint project of the University of California, Los Angeles, and the Jet Propulsion Laboratory/California Institute of Technology, and NEOWISE, which is a project of the Jet Propulsion Laboratory/California Institute of Technology. WISE and NEOWISE are funded by the National Aeronautics and Space Administration.

 The National Radio Astronomy Observatory is a facility of the National Science Foundation operated under cooperative agreement by Associated Universities, Inc.

This research made use of {\tt ds9}, a tool for data visualization supported by the Chandra X-ray Science Center (CXC) and the High Energy Astrophysics Science Archive Center (HEASARC) with support from the {\it JWST} Mission office at the Space Telescope Science Institute for 3D visualization. This paper also made use of the {\tt TOPCAT} software \citep{Taylor:05}.  Furthermore, this research has made use of the NASA/IPAC Extragalactic Database (NED), which is operated by the Jet Propulsion Laboratory, California Institute of Technology, under contract with the National Aeronautics and Space Administration. 

\end{acknowledgements}

\bibliographystyle{pasa-mnras}
\bibliography{grg_pasa}

\begin{thebibliography}{}
\makeatletter
\relax
\def\mn@urlcharsother{\let\do\@makeother \do\$\do\&\do\#\do\^\do\_\do\%\do\~}
\definecolor{darkblue}{rgb}{0,0,0.597656}
\def\mndoi{\begingroup\mn@urlcharsother \@ifnextchar [ {\mndoi@} {\mndoi@[]}}
\def\mndoi@[#1]#2{\def\@tempa{#1}\ifx\@tempa\@empty \href
  {http://dx.doi.org/#2} {\textcolor{darkblue}{doi:#2}}\else \href
  {http://dx.doi.org/#2} {\textcolor{darkblue}{#1}}\fi \endgroup}
\def\mn@eprint#1#2{\mn@eprint@#1:#2::\@nil}
\def\mn@eprint@arXiv#1{\href {http://arxiv.org/abs/#1} {{\tt arXiv:#1}}}
\def\mn@eprint@dblp#1{\href {http://dblp.uni-trier.de/rec/bibtex/#1.xml}
  {dblp:#1}}
\def\mn@eprint@#1:#2:#3:#4\@nil{\def\@tempa {#1}\def\@tempb {#2}\def\@tempc
  {#3}\ifx \@tempc \@empty \let \@tempc \@tempb \let \@tempb \@tempa \fi \ifx
  \@tempb \@empty \def\@tempb {arXiv}\fi \@ifundefined
  {mn@eprint@\@tempb}{\@tempb:\@tempc}{\expandafter \expandafter \csname
  mn@eprint@\@tempb\endcsname \expandafter{\@tempc}}}

\bibitem[\protect\citeauthoryear{{Abell}, {Corwin}  \& {Olowin}}{{Abell}
  et~al.}{1989}]{Abell:89}
{Abell} G.~O.,  {Corwin} Jr. H.~G.,   {Olowin} R.~P.,  1989, \mndoi [\apjs]
  {10.1086/191333}, \href {http://adsabs.harvard.edu/abs/1989ApJS...70....1A}
  {70, 1}

\bibitem[\protect\citeauthoryear{Akaike}{Akaike}{1974}]{akaike_new_1974}
Akaike H.,  1974, IEEE Transactions on Automatic Control, 19, 716

\bibitem[\protect\citeauthoryear{{Baars}, {Genzel}, {Pauliny-Toth}  \&
  {Witzel}}{{Baars} et~al.}{1977}]{Baars:77}
{Baars} J.~W.~M.,  {Genzel} R.,  {Pauliny-Toth} I.~I.~K.,   {Witzel} A.,  1977,
  \aap, \href {http://adsabs.harvard.edu/abs/1977A%26A....61...99B} {61, 99}

\bibitem[\protect\citeauthoryear{{Baldwin}, {Phillips}  \&
  {Terlevich}}{{Baldwin} et~al.}{1981}]{Baldwin:81}
{Baldwin} J.~A.,  {Phillips} M.~M.,   {Terlevich} R.,  1981, \mndoi [\pasp]
  {10.1086/130766}, \href {http://adsabs.harvard.edu/abs/1981PASP...93....5B}
  {93, 5}

\bibitem[\protect\citeauthoryear{{Battye} \& {Browne}}{{Battye} \&
  {Browne}}{2009}]{Battye:09}
{Battye} R.~A.,  {Browne} I.~W.~A.,  2009, \mndoi [\mnras]
  {10.1111/j.1365-2966.2009.15429.x}, \href
  {https://ui.adsabs.harvard.edu/abs/2009MNRAS.399.1888B} {399, 1888}

\bibitem[\protect\citeauthoryear{{Bock}, {Large}  \& {Sadler}}{{Bock}
  et~al.}{1999}]{Bock:99}
{Bock} D.~C.-J.,  {Large} M.~I.,   {Sadler} E.~M.,  1999, \mndoi [\aj]
  {10.1086/300786}, \href {http://adsabs.harvard.edu/abs/1999AJ....117.1578B}
  {117, 1578}

\bibitem[\protect\citeauthoryear{{B{\"o}hringer}, {Chon}, {Collins}, {Guzzo},
  {Nowak}  \& {Bobrovskyi}}{{B{\"o}hringer} et~al.}{2013}]{Bohringer:13}
{B{\"o}hringer} H.,  {Chon} G.,  {Collins} C.~A.,  {Guzzo} L.,  {Nowak} N.,
  {Bobrovskyi} S.,  2013, \mndoi [\aap] {10.1051/0004-6361/201220722}, \href
  {http://adsabs.harvard.edu/abs/2013A%26A...555A..30B} {555, A30}

\bibitem[\protect\citeauthoryear{{Brinchmann}, {Charlot}, {White}, {Tremonti},
  {Kauffmann}, {Heckman}  \& {Brinkmann}}{{Brinchmann}
  et~al.}{2004}]{Brinchmann:04}
{Brinchmann} J.,  {Charlot} S.,  {White} S.~D.~M.,  {Tremonti} C.,  {Kauffmann}
  G.,  {Heckman} T.,   {Brinkmann} J.,  2004, \mndoi [\mnras]
  {10.1111/j.1365-2966.2004.07881.x}, \href
  {http://adsabs.harvard.edu/abs/2004MNRAS.351.1151B} {351, 1151}

\bibitem[\protect\citeauthoryear{{Brown} \& {Burns}}{{Brown} \&
  {Burns}}{1991}]{Brown:91}
{Brown} D.~L.,  {Burns} J.~O.,  1991, \mndoi [\aj] {10.1086/116012}, \href
  {http://adsabs.harvard.edu/abs/1991AJ....102.1917B} {102, 1917}

\bibitem[\protect\citeauthoryear{{Brown}, {Jarrett}  \& {Cluver}}{{Brown}
  et~al.}{2014}]{Brown:14}
{Brown} M.~J.~I.,  {Jarrett} T.~H.,   {Cluver} M.~E.,  2014, \mndoi [\pasa]
  {10.1017/pasa.2014.44}, \href
  {http://adsabs.harvard.edu/abs/2014PASA...31...49B} {31, e049}

\bibitem[\protect\citeauthoryear{Burnham \& Anderson}{Burnham \&
  Anderson}{2002}]{burnham_model_2002}
Burnham K.~P.,  Anderson D.~R.,  2002, Model {Selection} and {Multimodel}
  {Inference}: {A} {Practical} {Information}-{Theoretic} {Approach}, 2 edn.
Springer-Verlag, New York, \url
  {https://www.springer.com/gp/book/9780387953649}

\bibitem[\protect\citeauthoryear{{Callingham} et~al.,}{{Callingham}
  et~al.}{2017}]{Callingham:17}
{Callingham} J.~R.,  et~al., 2017, \mndoi [\apj] {10.3847/1538-4357/836/2/174},
  \href {http://adsabs.harvard.edu/abs/2017ApJ...836..174C} {836, 174}

\bibitem[\protect\citeauthoryear{{Calzetti} et~al.,}{{Calzetti}
  et~al.}{2010}]{Calzetti:10}
{Calzetti} D.,  et~al., 2010, \mndoi [\apj] {10.1088/0004-637X/714/2/1256},
  \href {http://adsabs.harvard.edu/abs/2010ApJ...714.1256C} {714, 1256}

\bibitem[\protect\citeauthoryear{{Colless} et~al.,}{{Colless}
  et~al.}{2001}]{Colless:01}
{Colless} M.,  et~al., 2001, \mndoi [\mnras]
  {10.1046/j.1365-8711.2001.04902.x}, \href
  {http://adsabs.harvard.edu/abs/2001MNRAS.328.1039C} {328, 1039}

\bibitem[\protect\citeauthoryear{Condon, Cotton, Greisen, Yin, Perley, Perley,
  Taylor  \& Broderick}{Condon et~al.}{1998}]{Condon:98}
Condon J.~J.,  Cotton W.~D.,  Greisen E.~W.,  Yin Q.~F.,  Perley R.~A.,  Perley
  G.~B.,  Taylor G.~B.,   Broderick J.~J.,  1998, The Astronomical Journal,
  115, 1693

\bibitem[\protect\citeauthoryear{{Croston}, {Ineson}, {Hardcastle}  \&
  {Mingo}}{{Croston} et~al.}{2017}]{Croston:17}
{Croston} J.~H.,  {Ineson} J.,  {Hardcastle} M.~J.,   {Mingo} B.,  2017, \mndoi
  [\mnras] {10.1093/mnras/stx1347}, \href
  {http://adsabs.harvard.edu/abs/2017MNRAS.470.1943C} {470, 1943}

\bibitem[\protect\citeauthoryear{{Driver} et~al.,}{{Driver}
  et~al.}{2009}]{Driver:09}
{Driver} S.~P.,  et~al., 2009, \mndoi [Astronomy and Geophysics]
  {10.1111/j.1468-4004.2009.50512.x}, \href
  {http://adsabs.harvard.edu/abs/2009A%26G....50e..12D} {50, 5.12}

\bibitem[\protect\citeauthoryear{{Drouart} et~al.,}{{Drouart}
  et~al.}{2014}]{Drouart:14}
{Drouart} G.,  et~al., 2014, \aap, 566, A53

\bibitem[\protect\citeauthoryear{{Eales} et~al.}{{Eales}
  et~al.}{2010}]{Eales:10}
{Eales} S.,  et~al., 2010, preprint (\mn@eprint {arXiv} {1005.2189})

\bibitem[\protect\citeauthoryear{{Fabian} \& {Rees}}{{Fabian} \&
  {Rees}}{1995}]{Fabian:95}
{Fabian} A.~C.,  {Rees} M.~J.,  1995, \mndoi [\mnras]
  {10.1093/mnras/277.1.L55}, \href
  {http://adsabs.harvard.edu/abs/1995MNRAS.277L..55F} {277, L55}

\bibitem[\protect\citeauthoryear{{Fanaroff} \& {Riley}}{{Fanaroff} \&
  {Riley}}{1974}]{Fanaroff:74}
{Fanaroff} B.~L.,  {Riley} J.~M.,  1974, \mndoi [\mnras]
  {10.1093/mnras/167.1.31P}, \href
  {http://adsabs.harvard.edu/abs/1974MNRAS.167P..31F} {167, 31P}

\bibitem[\protect\citeauthoryear{{Fender}, {Belloni}  \& {Gallo}}{{Fender}
  et~al.}{2004}]{Fender:04}
{Fender} R.~P.,  {Belloni} T.~M.,   {Gallo} E.,  2004, \mndoi [\mnras]
  {10.1111/j.1365-2966.2004.08384.x}, \href
  {http://adsabs.harvard.edu/abs/2004MNRAS.355.1105F} {355, 1105}

\bibitem[\protect\citeauthoryear{Ferrarese \& Merrit}{Ferrarese \&
  Merrit}{2000}]{Ferrarese:00}
Ferrarese L.,  Merrit D.,  2000, ApJ, 539, L9

\bibitem[\protect\citeauthoryear{{Gaspari}, {Ruszkowski}  \& {Oh}}{{Gaspari}
  et~al.}{2013}]{Gaspari:13}
{Gaspari} M.,  {Ruszkowski} M.,   {Oh} S.~P.,  2013, \mndoi [\mnras]
  {10.1093/mnras/stt692}, \href
  {http://adsabs.harvard.edu/abs/2013MNRAS.432.3401G} {432, 3401}

\bibitem[\protect\citeauthoryear{{Godfrey} \& {Shabala}}{{Godfrey} \&
  {Shabala}}{2013}]{Godfrey:13}
{Godfrey} L.~E.~H.,  {Shabala} S.~S.,  2013, \mndoi [\apj]
  {10.1088/0004-637X/767/1/12}, \href
  {http://adsabs.harvard.edu/abs/2013ApJ...767...12G} {767, 12}

\bibitem[\protect\citeauthoryear{{Griffin} et~al.}{{Griffin}
  et~al.}{2010}]{Griffin:10}
{Griffin} M.~J.,  et~al., 2010, preprint (\mn@eprint {arXiv} {1005.5123})

\bibitem[\protect\citeauthoryear{{Hancock}, {Trott}  \&
  {Hurley-Walker}}{{Hancock} et~al.}{2018}]{Hancock:18}
{Hancock} P.~J.,  {Trott} C.~M.,   {Hurley-Walker} N.,  2018, \mndoi [\pasa]
  {10.1017/pasa.2018.3}, \href
  {https://ui.adsabs.harvard.edu/abs/2018PASA...35...11H} {35, e011}

\bibitem[\protect\citeauthoryear{{Hardcastle} \& {Croston}}{{Hardcastle} \&
  {Croston}}{2010}]{Hardcastle:10}
{Hardcastle} M.~J.,  {Croston} J.~H.,  2010, \mndoi [\mnras]
  {10.1111/j.1365-2966.2010.16420.x}, \href
  {http://adsabs.harvard.edu/abs/2010MNRAS.404.2018H} {404, 2018}

\bibitem[\protect\citeauthoryear{Hardcastle, Lawrence  \& Worrall}{Hardcastle
  et~al.}{1998}]{Hardcastle:98}
Hardcastle M.~J.,  Lawrence C.~R.,   Worrall D.~M.,  1998, ApJ, 504, 743

\bibitem[\protect\citeauthoryear{{Hardcastle} et~al.,}{{Hardcastle}
  et~al.}{2019}]{Hardcastle:19}
{Hardcastle} M.~J.,  et~al., 2019, \mndoi [\aap] {10.1051/0004-6361/201833893},
  \href {https://ui.adsabs.harvard.edu/abs/2019A&A...622A..12H} {622, A12}

\bibitem[\protect\citeauthoryear{{H{\"a}ring} \& {Rix}}{{H{\"a}ring} \&
  {Rix}}{2004}]{Haring:04}
{H{\"a}ring} N.,  {Rix} H.-W.,  2004, \mndoi [\apj] {10.1086/383567}, \href
  {https://ui.adsabs.harvard.edu/#abs/2004ApJ...604L..89H} {604, L89}

\bibitem[\protect\citeauthoryear{{Heavens} \& {Drury}}{{Heavens} \&
  {Drury}}{1988}]{Heavens:88}
{Heavens} A.~F.,  {Drury} L.~O.,  1988, \mndoi [\mnras]
  {10.1093/mnras/235.3.997}, \href
  {https://ui.adsabs.harvard.edu/abs/1988MNRAS.235..997H} {235, 997}

\bibitem[\protect\citeauthoryear{{Heckman} \& {Best}}{{Heckman} \&
  {Best}}{2014}]{Heckman:14}
{Heckman} T.~M.,  {Best} P.~N.,  2014, \mndoi [\araa]
  {10.1146/annurev-astro-081913-035722}, \href
  {http://adsabs.harvard.edu/abs/2014ARA%26A..52..589H} {52, 589}

\bibitem[\protect\citeauthoryear{{Heckman}, {Kauffmann}, {Brinchmann},
  {Charlot}, {Tremonti}  \& {White}}{{Heckman} et~al.}{2004}]{Heckman:04}
{Heckman} T.~M.,  {Kauffmann} G.,  {Brinchmann} J.,  {Charlot} S.,  {Tremonti}
  C.,   {White} S.~D.~M.,  2004, \mndoi [\apj] {10.1086/422872}, \href
  {http://adsabs.harvard.edu/abs/2004ApJ...613..109H} {613, 109}

\bibitem[\protect\citeauthoryear{{Hurley-Walker} et~al.,}{{Hurley-Walker}
  et~al.}{2017}]{NHW:17}
{Hurley-Walker} N.,  et~al., 2017, \mndoi [\mnras] {10.1093/mnras/stw2337},
  \href {http://adsabs.harvard.edu/abs/2017MNRAS.464.1146H} {464, 1146}

\bibitem[\protect\citeauthoryear{{Huynh}, {Bell}, {Hopkins}, {Norris}  \&
  {Seymour}}{{Huynh} et~al.}{2015}]{Huynh:15}
{Huynh} M.~T.,  {Bell} M.~E.,  {Hopkins} A.~M.,  {Norris} R.~P.,   {Seymour}
  N.,  2015, \mndoi [\mnras] {10.1093/mnras/stv2012}, \href
  {https://ui.adsabs.harvard.edu/abs/2015MNRAS.454..952H} {454, 952}

\bibitem[\protect\citeauthoryear{{Intema}, {van der Tol}, {Cotton}, {Cohen},
  {van Bemmel}  \& {R{\"o}ttgering}}{{Intema} et~al.}{2009}]{Intema:09}
{Intema} H.~T.,  {van der Tol} S.,  {Cotton} W.~D.,  {Cohen} A.~S.,  {van
  Bemmel} I.~M.,   {R{\"o}ttgering} H.~J.~A.,  2009, \mndoi [\aap]
  {10.1051/0004-6361/200811094}, \href
  {http://adsabs.harvard.edu/abs/2009A%26A...501.1185I} {501, 1185}

\bibitem[\protect\citeauthoryear{{Intema}, {Jagannathan}, {Mooley}  \&
  {Frail}}{{Intema} et~al.}{2017}]{Intema:17}
{Intema} H.~T.,  {Jagannathan} P.,  {Mooley} K.~P.,   {Frail} D.~A.,  2017,
  \mndoi [\aap] {10.1051/0004-6361/201628536}, \href
  {http://adsabs.harvard.edu/abs/2017A%26A...598A..78I} {598, A78}

\bibitem[\protect\citeauthoryear{{Ishwara-Chandra} \&
  {Saikia}}{{Ishwara-Chandra} \& {Saikia}}{1999}]{Ishwar:99}
{Ishwara-Chandra} C.~H.,  {Saikia} D.~J.,  1999, \mndoi [\mnras]
  {10.1046/j.1365-8711.1999.02835.x}, \href
  {http://adsabs.harvard.edu/abs/1999MNRAS.309..100I} {309, 100}

\bibitem[\protect\citeauthoryear{{Jarrett} et~al.,}{{Jarrett}
  et~al.}{2011}]{Jarrett:11}
{Jarrett} T.~H.,  et~al., 2011, \mndoi [\apj] {10.1088/0004-637X/735/2/112},
  \href {http://adsabs.harvard.edu/abs/2011ApJ...735..112J} {735, 112}

\bibitem[\protect\citeauthoryear{{Johnston} et~al.,}{{Johnston}
  et~al.}{2007}]{Johnston:07}
{Johnston} S.,  et~al., 2007, \mndoi [Publications of the Astronomical Society
  of Australia] {10.1071/AS07033}, \href
  {https://ui.adsabs.harvard.edu/\#abs/2007PASA...24..174J} {24, 174}

\bibitem[\protect\citeauthoryear{{Komberg} \& {Pashchenko}}{{Komberg} \&
  {Pashchenko}}{2009}]{Komberg:09}
{Komberg} B.~V.,  {Pashchenko} I.~N.,  2009, \mndoi [Astronomy Reports]
  {10.1134/S1063772909120026}, \href
  {https://ui.adsabs.harvard.edu/abs/2009ARep...53.1086K} {53, 1086}

\bibitem[\protect\citeauthoryear{{Kormendy} \& {Ho}}{{Kormendy} \&
  {Ho}}{2013}]{Kormendy:13}
{Kormendy} J.,  {Ho} L.~C.,  2013, \mndoi [\araa]
  {10.1146/annurev-astro-082708-101811}, \href
  {http://adsabs.harvard.edu/abs/2013ARA%26A..51..511K} {51, 511}

\bibitem[\protect\citeauthoryear{{Krause} et~al.,}{{Krause}
  et~al.}{2019}]{Krause:19}
{Krause} M.~G.~H.,  et~al., 2019, \mndoi [\mnras] {10.1093/mnras/sty2558},
  \href {http://adsabs.harvard.edu/abs/2019MNRAS.482..240K} {482, 240}

\bibitem[\protect\citeauthoryear{{Ku{\'z}micz}, {Jamrozy}, {Bronarska},
  {Janda-Boczar}  \& {Saikia}}{{Ku{\'z}micz} et~al.}{2018}]{Kuzmicz:18}
{Ku{\'z}micz} A.,  {Jamrozy} M.,  {Bronarska} K.,  {Janda-Boczar} K.,
  {Saikia} D.~J.,  2018, \mndoi [\apjs] {10.3847/1538-4365/aad9ff}, \href
  {https://ui.adsabs.harvard.edu/abs/2018ApJS..238....9K} {238, 9}

\bibitem[\protect\citeauthoryear{{Lacy}, {Ridgway}, {Sajina}, {Petric},
  {Gates}, {Urrutia}  \& {Storrie-Lombardi}}{{Lacy} et~al.}{2015}]{Lacy:15}
{Lacy} M.,  {Ridgway} S.~E.,  {Sajina} A.,  {Petric} A.~O.,  {Gates} E.~L.,
  {Urrutia} T.,   {Storrie-Lombardi} L.~J.,  2015, \mndoi [\apj]
  {10.1088/0004-637X/802/2/102}, \href
  {http://adsabs.harvard.edu/abs/2015ApJ...802..102L} {802, 102}

\bibitem[\protect\citeauthoryear{{Leahy} et~al.,}{{Leahy}
  et~al.}{2019}]{Leahy:19}
{Leahy} D.~A.,  et~al., 2019, \mndoi [\pasa] {10.1017/pasa.2019.16}, \href
  {https://ui.adsabs.harvard.edu/abs/2019PASA...36...24L} {36, e024}

\bibitem[\protect\citeauthoryear{{Mateos} et~al.,}{{Mateos}
  et~al.}{2012}]{Mateos:12}
{Mateos} S.,  et~al., 2012, \mnras, 426, 3271

\bibitem[\protect\citeauthoryear{{McConnell} et~al.,}{{McConnell}
  et~al.}{2016}]{McConnell:16}
{McConnell} D.,  et~al., 2016, \mndoi [\pasa] {10.1017/pasa.2016.37}, \href
  {http://adsabs.harvard.edu/abs/2016PASA...33...42M} {33, e042}

\bibitem[\protect\citeauthoryear{{McMullin}, {Waters}, {Schiebel}, {Young}  \&
  {Golap}}{{McMullin} et~al.}{2007}]{McMullin:07}
{McMullin} J.~P.,  {Waters} B.,  {Schiebel} D.,  {Young} W.,   {Golap} K.,
  2007, in {Shaw} R.~A.,  {Hill} F.,   {Bell} D.~J.,  eds,  Astronomical
  Society of the Pacific Conference Series Vol. 376, Astronomical Data Analysis
  Software and Systems XVI. p.~127

\bibitem[\protect\citeauthoryear{{Meier}}{{Meier}}{2002}]{Meier:02}
{Meier} D.~L.,  2002, \mndoi [New Astronomy Reviews]
  {10.1016/S1387-6473(01)00189-0}, \href
  {http://adsabs.harvard.edu/abs/2002NewAR..46..247M} {46, 247}

\bibitem[\protect\citeauthoryear{{Merloni}, {Heinz}  \& {di Matteo}}{{Merloni}
  et~al.}{2003}]{Merloni:03}
{Merloni} A.,  {Heinz} S.,   {di Matteo} T.,  2003, \mndoi [\mnras]
  {10.1046/j.1365-2966.2003.07017.x}, \href
  {http://adsabs.harvard.edu/abs/2003MNRAS.345.1057M} {345, 1057}

\bibitem[\protect\citeauthoryear{{M$^{\rm c}$Hardy}, {Koerding}, {Knigge},
  {Uttley}  \& {Fender}}{{M$^{\rm c}$Hardy} et~al.}{2006}]{McHardy:06}
{M$^{\rm c}$Hardy} I.~M.,  {Koerding} E.,  {Knigge} C.,  {Uttley} P.,
  {Fender} R.~P.,  2006, \mndoi [\nat] {10.1038/nature05389}, \href
  {http://adsabs.harvard.edu/abs/2006Natur.444..730M} {444, 730}

\bibitem[\protect\citeauthoryear{{Murgia} et~al.,}{{Murgia}
  et~al.}{2011}]{Murgia:11}
{Murgia} M.,  et~al., 2011, \mndoi [\aap] {10.1051/0004-6361/201015302}, \href
  {https://ui.adsabs.harvard.edu/abs/2011A%26A...526A.148M} {526, A148}

\bibitem[\protect\citeauthoryear{{Murphy} et~al.}{{Murphy}
  et~al.}{2006}]{Murphy:06}
{Murphy} E.~J.,  et~al., 2006, \apjl, 651, L111

\bibitem[\protect\citeauthoryear{{Murphy} et~al.,}{{Murphy}
  et~al.}{2010}]{Murphy:10}
{Murphy} T.,  et~al., 2010, \mndoi [\mnras] {10.1111/j.1365-2966.2009.15961.x},
  \href {http://adsabs.harvard.edu/abs/2010MNRAS.402.2403M} {402, 2403}

\bibitem[\protect\citeauthoryear{{Noeske} et~al.,}{{Noeske}
  et~al.}{2007}]{Noeske:07}
{Noeske} K.~G.,  et~al., 2007, \mndoi [\apjl] {10.1086/517926}, \href
  {http://adsabs.harvard.edu/abs/2007ApJ...660L..43N} {660, L43}

\bibitem[\protect\citeauthoryear{{Norris} et~al.,}{{Norris}
  et~al.}{2011}]{Norris:11}
{Norris} R.~P.,  et~al., 2011, \mndoi [\pasa] {10.1071/AS11021}, \href
  {http://adsabs.harvard.edu/abs/2011PASA...28..215N} {28, 215}

\bibitem[\protect\citeauthoryear{{Novak}, {Ostriker}  \& {Ciotti}}{{Novak}
  et~al.}{2011}]{Novak:11}
{Novak} G.~S.,  {Ostriker} J.~P.,   {Ciotti} L.,  2011, \mndoi [\apj]
  {10.1088/0004-637X/737/1/26}, \href
  {http://adsabs.harvard.edu/abs/2011ApJ...737...26N} {737, 26}

\bibitem[\protect\citeauthoryear{Partridge, L{\'{o}}pez-Caniego, Perley,
  Stevens, Butler, Rocha, Walter  \& Zacchei}{Partridge
  et~al.}{2016}]{Partridge:16}
Partridge B.,  L{\'{o}}pez-Caniego M.,  Perley R.~A.,  Stevens J.,  Butler
  B.~J.,  Rocha G.,  Walter B.,   Zacchei A.,  2016, \mndoi [The Astrophysical
  Journal] {10.3847/0004-637x/821/1/61}, 821, 61

\bibitem[\protect\citeauthoryear{{Poglitsch} et~al.,}{{Poglitsch}
  et~al.}{2010}]{Poglitsch:10}
{Poglitsch} A.,  et~al., 2010, \aap, 518, L2

\bibitem[\protect\citeauthoryear{Rees}{Rees}{1978}]{Rees:78}
Rees M.~J.,  1978, Nature, 275, 516

\bibitem[\protect\citeauthoryear{Reynolds}{Reynolds}{1994}]{Reynolds_1994}
Reynolds J.~E.,  1994, Technical report, {A Revised Flux Scale for the AT
  Compact Array}.
ATNF, Epping

\bibitem[\protect\citeauthoryear{{Robotham} et~al.,}{{Robotham}
  et~al.}{2011}]{Robotham:11}
{Robotham} A.~S.~G.,  et~al., 2011, \mndoi [\mnras]
  {10.1111/j.1365-2966.2011.19217.x}, \href
  {http://adsabs.harvard.edu/abs/2011MNRAS.416.2640R} {416, 2640}

\bibitem[\protect\citeauthoryear{{Robotham}, {Davies}, {Driver}, {Koushan},
  {Taranu}, {Casura}  \& {Liske}}{{Robotham} et~al.}{2018}]{Robotham:18}
{Robotham} A.~S.~G.,  {Davies} L.~J.~M.,  {Driver} S.~P.,  {Koushan} S.,
  {Taranu} D.~S.,  {Casura} S.,   {Liske} J.,  2018, \mndoi [\mnras]
  {10.1093/mnras/sty440}, \href
  {http://adsabs.harvard.edu/abs/2018MNRAS.476.3137R} {476, 3137}

\bibitem[\protect\citeauthoryear{{Sarazin}}{{Sarazin}}{1988}]{Sarazin:88}
{Sarazin} C.~L.,  1988, {X-ray emission from clusters of galaxies}.
Cambridge Astrophysics Series, Cambridge: Cambridge University Press

\bibitem[\protect\citeauthoryear{{Sault}, {Teuben}  \& {Wright}}{{Sault}
  et~al.}{1995}]{Sault:95}
{Sault} R.~J.,  {Teuben} P.~J.,   {Wright} M.~C.~H.,  1995, in {Shaw} R.~A.,
  {Payne} H.~E.,   {Hayes} J.~J.~E.,  eds,  Astronomical Society of the Pacific
  Conference Series Vol. 77, Astronomical Data Analysis Software and Systems
  IV. p.~433 (\mn@eprint {} {astro-ph/0612759})

\bibitem[\protect\citeauthoryear{{Seymour} et~al.}{{Seymour}
  et~al.}{2008}]{Seymour:08}
{Seymour} N.,  et~al., 2008, \mnras, 386, 1695

\bibitem[\protect\citeauthoryear{{Shabala} \& {Godfrey}}{{Shabala} \&
  {Godfrey}}{2013}]{Shabala:13}
{Shabala} S.~S.,  {Godfrey} L.~E.~H.,  2013, \mndoi [\apj]
  {10.1088/0004-637X/769/2/129}, \href
  {http://adsabs.harvard.edu/abs/2013ApJ...769..129S} {769, 129}

\bibitem[\protect\citeauthoryear{{Shabala}, {Ash}, {Alexander}  \&
  {Riley}}{{Shabala} et~al.}{2008}]{Shabala:08}
{Shabala} S.~S.,  {Ash} S.,  {Alexander} P.,   {Riley} J.~M.,  2008, \mndoi
  [\mnras] {10.1111/j.1365-2966.2008.13459.x}, \href
  {https://ui.adsabs.harvard.edu/abs/2008MNRAS.388..625S} {388, 625}

\bibitem[\protect\citeauthoryear{{Shabala}, {Deller}, {Kaviraj}, {Middelberg},
  {Turner}, {Ting}, {Allison}  \& {Davis}}{{Shabala} et~al.}{2017}]{Shabala:17}
{Shabala} S.~S.,  {Deller} A.,  {Kaviraj} S.,  {Middelberg} E.,  {Turner}
  R.~J.,  {Ting} Y.~S.,  {Allison} J.~R.,   {Davis} T.~A.,  2017, \mndoi
  [\mnras] {10.1093/mnras/stw2536}, \href
  {http://adsabs.harvard.edu/abs/2017MNRAS.464.4706S} {464, 4706}

\bibitem[\protect\citeauthoryear{Skrutskie et~al.}{Skrutskie
  et~al.}{2006}]{Skrutskie:06}
Skrutskie M.~F.,  et~al., 2006, The Astronomical Journal, 131, 1163

\bibitem[\protect\citeauthoryear{{Symeonidis}, {Giblin}, {Page}, {Pearson},
  {Bendo}, {Seymour}  \& {Oliver}}{{Symeonidis} et~al.}{2016}]{Symeonidis:16}
{Symeonidis} M.,  {Giblin} B.~M.,  {Page} M.~J.,  {Pearson} C.,  {Bendo} G.,
  {Seymour} N.,   {Oliver} S.~J.,  2016, \mnras, 459, 257

\bibitem[\protect\citeauthoryear{{Taylor}}{{Taylor}}{2005}]{Taylor:05}
{Taylor} M.~B.,  2005, in {Shopbell} P.,  {Britton} M.,   {Ebert} R.,  eds,
  Astronomical Society of the Pacific Conference Series Vol. 347, Astronomical
  Data Analysis Software and Systems XIV. p.~29

\bibitem[\protect\citeauthoryear{{Tingay} et~al.,}{{Tingay}
  et~al.}{2013}]{Tingay:13}
{Tingay} S.~J.,  et~al., 2013, \mndoi [\pasa] {10.1017/pasa.2012.007}, \href
  {http://adsabs.harvard.edu/abs/2013PASA...30....7T} {30, e007}

\bibitem[\protect\citeauthoryear{{Turner} \& {Shabala}}{{Turner} \&
  {Shabala}}{2015}]{Turner:15}
{Turner} R.~J.,  {Shabala} S.~S.,  2015, \mndoi [\apj]
  {10.1088/0004-637X/806/1/59}, \href
  {http://adsabs.harvard.edu/abs/2015ApJ...806...59T} {806, 59}

\bibitem[\protect\citeauthoryear{{Turner}, {Rogers}, {Shabala}  \&
  {Krause}}{{Turner} et~al.}{2018a}]{Turner:18a}
{Turner} R.~J.,  {Rogers} J.~G.,  {Shabala} S.~S.,   {Krause} M.~G.~H.,  2018a,
  \mndoi [\mnras] {10.1093/mnras/stx2591}, \href
  {http://adsabs.harvard.edu/abs/2018MNRAS.473.4179T} {473, 4179}

\bibitem[\protect\citeauthoryear{{Turner}, {Shabala}  \& {Krause}}{{Turner}
  et~al.}{2018b}]{Turner:18b}
{Turner} R.~J.,  {Shabala} S.~S.,   {Krause} M.~G.~H.,  2018b, \mndoi [\mnras]
  {10.1093/mnras/stx2947}, \href
  {http://adsabs.harvard.edu/abs/2018MNRAS.474.3361T} {474, 3361}

\bibitem[\protect\citeauthoryear{Urry \& Padovani}{Urry \&
  Padovani}{1995}]{Urry:95}
Urry C.~M.,  Padovani P.,  1995, PASP, 107, 803

\bibitem[\protect\citeauthoryear{Voges et~al.}{Voges et~al.}{1999}]{Voges:99}
Voges W.,  et~al., 1999, Astronomy \& Astrophysics, 349, 389

\bibitem[\protect\citeauthoryear{{Wayth} et~al.,}{{Wayth}
  et~al.}{2015}]{wayth:15}
{Wayth} R.~B.,  et~al., 2015, \mndoi [\pasa] {10.1017/pasa.2015.26}, \href
  {http://adsabs.harvard.edu/abs/2015PASA...32...25W} {32, e025}

\bibitem[\protect\citeauthoryear{{Wright}}{{Wright}}{2006}]{Wright:06}
{Wright} E.~L.,  2006, \pasp, 118, 1711

\bibitem[\protect\citeauthoryear{{Wright} et~al.,}{{Wright}
  et~al.}{2010}]{Wright:10}
{Wright} E.~L.,  et~al., 2010, \mndoi [\aj] {10.1088/0004-6256/140/6/1868},
  \href {http://adsabs.harvard.edu/abs/2010AJ....140.1868W} {140, 1868}

\makeatother
\end{thebibliography}

\end{document}